\documentstyle[12pt,aaspp4]{article} 
\def\tempest%
{\begin{array}{ccc} 
1 & 1 & 1 \\ 
1 & 1 & 1 \\ 
4 & 3 & 8 
\end{array}}

\def\bmu{\hbox{$\mu\hskip-7.5pt\mu$}}
\def\kms{{\rm km}\,{\rm s}^{-1}}
\def\masyr{{\rm mas}\, {\rm yr}^{-1}}
\begin{document}

\title{Nearby Microlensing Events - Identification of the Candidates for 
the Space Interferometry Mission}
\author 
{Samir Salim
and
Andrew Gould}
\affil{Ohio State University, Department of Astronomy, Columbus, OH 43210} 
\affil{E-mail: samir@astronomy.ohio-state.edu, gould@astronomy.ohio-state.edu} 
\begin{abstract} 
	The {\it Space Interferometry Mission (SIM)} is the instrument of choice when it comes to observing astrometric microlensing events where nearby, usually high-proper-motion stars (``lenses''), pass in front of more distant stars (``sources''). Each such encounter produces a deflection in the source's apparent position that when observed by {\it SIM} can lead to a precise mass determination of the nearby lens star.  We search for lens-source encounters during the 2005-2015 period using Hipparcos, ACT and NLTT to select lenses, and USNO-A2.0 to search for the corresponding sources, and rank these by the {\it SIM} time required for a 1\% mass measurement. 

	For Hipparcos and ACT lenses, the lens distance and lens-source impact parameter are precisely determined so the events are well characterized. We present 32 candidates beginning with a 61 Cyg A event in 2012 that requires only a few minutes of {\it SIM} time. Proxima Centauri and Barnard's star each generate several events. For NLTT lenses, the distance is known only to a factor of 3, and the impact parameter  only to $1''$. Together, these produce uncertainties of a factor $\sim 10$ in the amount of {\it SIM} time required. We present a list of 146 NLTT candidates and show how single-epoch CCD photometry of the candidates could reduce the uncertainty in {\it SIM} time to a factor of $\sim 1.5$.    
\keywords{astrometry -- Galaxy: stellar content -- gravitational lensing
-- stars: fundamental parameters (masses)} 
\end{abstract} 
\newpage

\section{Introduction} 
	
	One is used to thinking of microlensing events as taking place 
towards the Magellanic Clouds or the Galactic bulge. In both of these cases 
the lens is a faraway object, either a star belonging to the same system as 
the source star (self-lensing), a distant star in the Milky Way's disk, or a member of Milky Way's halo (whatever its nature might 
be). The effect that is routinely observed in such cases is the change in 
source's brightness, but also present is an additional effect of the deflection of the source
apparent position (Boden, Shao \& Van Buren 1998). The deflection is so small ($\sim 100 \mu$as) as to be unobservable with present-day facilities. However the unprecedented astrometric precision of the {\it Space Interferometry Mission (SIM)} ($4 \mu$as) will enable such 
measurements.

	{\it SIM} will also make possible observing microlensing events 
produced by {\it nearby} stars (``lenses'') moving in front of more distant stars (`` sources''). 
In such cases the deflection is the only observable effect, and so the encounters are referred to as {\it astrometric} microlensing events. Astrometry of such an event
 would yield the {\it mass} of the lens, i.e., the nearby star. Indeed this is the only known method to obtain the masses of stars not residing in binary systems. 
Initially proposed by Refsdal (1964), this idea was later examined by 
Paczy\'nski (1995, 1998) and Miralda-Escud\'e (1996) in the context of rapid developments in space-based astrometry. Gould (2000) quantified the {\it SIM} time required to achieve a given precision of mass measurement (say $\sigma_M / M = 1\%$) and investigated how the number of such measurements that could be made in a fixed {\it SIM} time depends on the characteristics of the search catalogs. The types of stars whose mass can be determined 
in this way will be discussed in \S\ 5.

	This paper can be seen as a direct answer to the question posed in 
\S\ 4 of Gould (2000) about what can be done to identify candidates using existing catalogs. We implement the suggestions
 given in that section (many attributable originally to I. Reid 1999, private communication) and expand on them to produce more accurate predictions. 
Specifically, we search for candidate events where a lens from a proper motion catalog [Hipparcos (ESA 1997), ACT (Urban et al. 1998b), NLTT  (Luyten 1979, 1980, Luyten \& Hughes 1980)] passes in front of a source from the USNO-A2.0 (Monet 1998) catalog.
 The features of these catalogs pertinent to candidate identification are 
discussed in the following section. Section 3 is devoted to a treatment 
of the estimate of the {\it SIM} time required for any of these candidate 
astrometric microlensing events in light of the 
limitations of the catalogs, in particular the absence of direct distance information 
in most cases. In \S\ 4 we discuss how we conducted the search for events and the extent to which we were able to overcome the various problems we encountered. In 
section \S\ 5 we present lists of candidate events that we decided are real.
 Positive identification on either DSS or paper edition of POSS I was done for all Hipparcos/ACT events.
 We also discuss the features of these events in general.

\section{Characteristics of the Catalogs}

	Our overall plan is to search for astrometric microlensing events (or 
`events', for short) and rank these by the amount of {\it SIM} time required to 
measure the lens mass to a fixed fractional error of 1\%.  To this end, we 
would like to consult a catalog containing the positions, parallaxes, proper 
motions, and magnitudes of all stellar sources in the sky.  Unfortunately, 
there is no such catalog.  To understand how to make use of existing catalogs,
 we review the basic requirements of the search.

	First, while in principle the event depends on the relative proper
motion of the source and lens, the lenses, being closer, almost always move  
much faster in the sky than
the sources.  Hence, no proper-motion information is required for the sources
in order to select candidate events. The USNO-A2.0 all sky astrometric catalog 
, which is constructed from two photographic surveys
[Palomar Observatory Sky Survey I (POSS I) for $\delta >-17.\hskip-2pt^\circ 5$ (`north' 
celestial hemisphere) and UK Science Research Council SRC-J survey plates and 
European Southern Observatory ESO-R survey plates (SERC/ESO) for $\delta 
<-17.\hskip-2pt^\circ 5$ (`south' celestial hemisphere)]
is therefore a nearly ideal catalog for sources, containing 526 million
 entries.  To be included in the catalog, a star had to be detected on both the
 blue and red plates within a $2''$ coincidence radius aperture.  Hence the 
catalog begins to lose completeness at
$V\sim 19$ as stars fall below the detection threshold on one plate or the
other.  The catalog is also incomplete at bright magnitudes $(V\la 11)$
because of poor astrometry of saturated stars, although for these stars 
USNO-A2.0 contains inserted entries from the ACT or Tycho (ESA 1997) 
catalog.  However, the epoch of these additional entries is 2000.0 and 1991.25
 respectively, unlike the epoch of the other sources which is the mean epoch 
of the blue and the red plates (1950s for POSS I, and 1980s for SERC/ESO).  In 
addition,  USNO-A2.0 is by and large missing the stars with proper motions $\mu \ga
 250\,\masyr$ in the `south',
because the blue and red plates of the SERC/ESO survey were on average taken 8
years apart, and so stars with $\mu>250\,\masyr$ moved
outside the $2''$ error circle between the blue and red exposures. In 
reality, the time elapsed between the two plates varies from 0 to 15 years, 
leading to different proper-motion cutoffs for each plate.
This problem does not affect POSS I because its blue and red plates were
taken on the same night.  Neither the incompleteness at bright magnitudes
nor the incompleteness at high proper motions has any significant effect on
USNO-A2.0 as a catalog for microlensing {\it sources}, since they are usually 
faint and move very slowly.  However, both have substantial impact
on our efforts to obtain critical information from this catalog 
about the {\it lenses} (see below).

	The {\it relative} position errors, important for NLTT events, for 
USNO-A2.0 are about 150 mas. 
For Hipparcos and ACT events, it is the absolute errors [USNO-A2.0 uses ICRS (International Celestial Reference System) as its reference frame] of about 250 mas that are relevant.   
There is, of course, an additional error in the position of the {\it source} in
2010 due to 60 years of proper motion in the case of POSS I and 30 years for 
the SERC/ESO plates.  Since typical
sources are on average 3 kpc distant and are moving at $25\,\kms$ in each 
direction, this gives a proper motion of $\sim 2\,\masyr$. This proper motion 
adds about 100 mas in the `north' and 50 mas in the `south' to the total 
positional error.
Hence, the total error on average is about 170 mas (260 mas in the absolute system).  Note that this will not be improved significantly
by the release of the USNO-B 
all-sky position and proper motion catalog
(D. Monet 1998, private communication), since its proper-motion errors
will be of the same order as the proper motions of typical source stars. Similar limitations will hold true in the case of GSC II (Guide Star Catalog Two) (see Morrison \& McLean 1999, for example). 
USNO-B and GSC II will be compiled by comparing first generation sky surveys with the 
second generation.  
The absolute photometry errors in USNO-A2.0 are said to be about 0.25 mag for 
the stars that are not saturated. USNO-A2.0 lists photographic blue and red 
magnitudes. The equinox of the coordinates is ICRS J2000.

	The probability $p$ that any individual lens will deflect light from
a more distant star enough to measure the lens mass $M$ to fixed fractional accuracy is 
\begin{equation}
p\propto N_s\pi\mu M,
\label{eqn:probscaling}
\end{equation}
where $N_s$ is the surface density of sources, and $\pi$ and $\mu$ are 
the parallax and proper motion of the lens.  One therefore expects events to
be clustered near the Galactic plane, and for nearby, fast-moving stars
to be over-represented as lenses.  However, there are a greater number of
 distant than
nearby stars and consequently more stars with low than high proper motions.
The net of these two competing effects is that for parallax-limited and
proper-motion-limited catalogs, the total number of events scales as
(Gould 2000)
\begin{equation}
N_{\rm events}\propto \pi_{\rm min}^{-1}, \qquad
N_{\rm events}\propto \mu_{\rm min}^{-1},
\label{eqn:pimulims}
\end{equation}
where $\pi_{\rm min}$ and $\mu_{\rm min}$ are the limits of the respective
types of catalogs of lenses.  Of course, the total number of potential lenses 
that one must examine scales as $\pi_{\rm min}^{-3}$ or $\mu_{\rm min}^{-3}$.
Thus it is most efficient to start with high $\pi$ or high $\mu$ stars and
move progressively to more distant or slower ones.  In practice, one has
available magnitude-limited and not distance-limited catalogs, but for stars
of fixed absolute magnitude these are effectively distance-limited.

	We search for lenses in three catalogs: Hipparcos, the `ACT Reference Catalog' (ACT) and the `New Luyten Catalogue of Stars with Proper Motions Larger than Two Tenths of an Arcsecond and First Supplement' (NLTT).
The three catalogs have substantially different characteristics.

	Hipparcos is a heterogeneous catalog with 118,000 entries.  However,
it has two approximate completeness characteristics that are 
very useful for understanding its role in the present study.  First, it is
approximately complete for $V<8$, with 41,000 stars to this limit.  Second,
it contains essentially all the NLTT stars brighter than its operational
limit of $V\sim 12$.  As we mention below, NLTT is nominally complete 
for $\mu>180\,\masyr$. Based on statistical
tests of the Hipparcos catalog, we find that it (and thus presumably NLTT)
is essentially complete for $\mu>220\,\masyr$ and $V<11$.  In its last 
magnitude $(11<V<12)$, Hipparcos 
shows some evidence for incompleteness, perhaps because of the
difficulty of making precise conversions from NLTT's photographic magnitudes
to the near-Johnson system used by Hipparcos.  There are 6500 Hipparcos
stars with  $\mu>200\,\masyr$ and 15,000 with $\mu>100\,\masyr$.

	Hipparcos stars have trigonometric parallaxes with typical
precisions of 1 mas.  As we discuss in \S\ 3, uncertainty in the distance
to the lens is the main problem in estimating the amount of {\it SIM} time 
required
for a lens mass measurement.  This uncertainty is virtually eliminated for
Hipparcos stars.  In addition, we use Hipparcos parallaxes to calibrate
our method for estimating distances of stars in the other two catalogs which
lack trigonometric parallaxes.  Hipparcos positions are accurate to 1 mas, 
while the proper motions have errors of
order $1\,\masyr$, implying an error of about 20 mas in the star's 2010
position.  This is negligible compared to the error in the source position
given in USNO-A2.0.  Finally, most Hipparcos stars have Tycho
photometry which is accurate to of order 0.01 mag.  Even those stars lacking
Tycho photometry usually have ground-based photometry of similar quality.
Tycho photometry is far better than the minimum precision required for the
present search.

	The ACT catalog is constructed by matching stars common to both 
the Astrographic Catalogue 2000 (AC 2000, Urban et al. 1998a) and Tycho, with epochs 
circa 1910 and 1990 respectively.  Such a long baseline combined with Tycho's precise 
positions, permits the proper motion accuracy of ACT to be $\sim 3\,\masyr$ (ten 
times better than Tycho itself). ACT 
is presently the largest (nearly 1 million stars) all-sky catalog containing proper 
motions. It is limited at the faint end by incompleteness
of the Tycho catalog which sets in over the range $11\la V\la 11.5$, and
at the bright end by incompleteness (due to saturation) of plates that produce AC 2000. 
Completeness of ACT with respect to Tycho (entries that have proper motion) is
 about 95\% in the $6<V<12$ range, 
and drops to 50\% for $V \sim 3$.  There is also a cutoff at high proper motions ($\mu\ga 1.\hskip-2pt''5\,{\rm yr}^{-1}$), which results from the lack of proper-motion information about these stars in the Tycho catalog. Typical errors of ACT proper motions
 imply an uncertainty in 2010 position of about 60 mas.
This is still small compared to the uncertainty of the source position and
so can be ignored.  Tycho photometry is available for the great majority
of ACT stars and, as stated above, this has much higher precision than
is required for the present study. 
As we discuss in \S\ 5.1, we are able to estimate the distances to ACT stars
with $\sim 30\%$ accuracy which is quite adequate for our purposes.

	NLTT is nominally complete to $\mu>180\,\masyr$
and $V<19$ in the northern part of the sky ($\delta \ga -33^\circ$), and away from the galactic plane ($\mid b\mid>10^{\circ}$) (see discussion in \S\ 5.). In the south and near the plane, the incompleteness sets in at brighter magnitudes.  NLTT $\alpha$ and $\delta$ 
are given only to 1 s and $0.\hskip-2pt'1$ respectively (in some cases to 0.1 
min and $1'$ respectively) and so are not
sufficiently accurate to predict lens-source encounters which typically
have impact parameters $\beta\sim 1''$.  Hence to obtain improved positions of NLTT stars we 
search for the corresponding entries in USNO-A2.0.  Recall that USNO-A2.0 entries have position errors of 250 mas.  However, recall also that in the `south' ($\delta<-17.\hskip-2pt^\circ 5$),
USNO-A2.0 is missing a large fraction of the NLTT stars.  
To recover this part of the 
NLTT catalog, it would be necessary to make new position measurements for the 
majority of NLTT stars
in the `south', or at least for all that pass within $6''$ (position error of NLTT) of some source star.
This would be a major project which we do not attempt.  Reid (1990) finds
that proper motion errors in NLTT are typically $20\,\masyr$ at the faint end.
By comparing NLTT and Hipparcos, we find a similar value at the bright end.
When this is propagated over a 60 year baseline (the epoch of NLTT is circa 
1950), it implies errors of
$1.\hskip-2pt ''2$ in 2010 position.  This is the dominant astrometric
error for these stars and has important consequences as we discuss in
\S\ 3.2 below.  

	Because NLTT stars must be found in USNO-A2.0 in order to
be used, they automatically have available two sources of photometry,
both photographic.  As we discuss in \S\ 3.1, it is necessary to
transform these photographic systems to the Johnson-like system used
by Tycho in order to estimate distances.  We find that the transformation
from USNO-A2.0 colors to Johnson $B-V$ has somewhat smaller scatter than
the transformation from NLTT colors, and we therefore use the former.  This
scatter (0.25 mag) is still substantially larger than we would like.  As
we discuss in \S\ 3.1, it leads to a factor 1.7 uncertainty in distance
estimates for NLTT stars.

	In brief, Hipparcos alone may be roughly thought of as complete for 
$V<8$ and for $\mu>180\,\masyr$ and $V<12$.  ACT is roughly complete for
$V<12$.  Thus, the combination of Hipparcos and ACT (which both have
high quality proper motions and distance estimates) is approximately
complete for $V<12$.  This sample is complemented by NLTT which is roughly
complete for $V<19$ and $\mu>180\,\masyr$, but has much lower quality
proper motion and distance estimates.

	There is one important additional source of incompleteness that affects all searches based on USNO-A2.0.  Suppose
that a lens will pass close to a source in 2010 with a relative proper motion
$\mu$.  The source must be identified from USNO-A2.0 which is based on
plates taken $\delta t\sim60\,$yr earlier in the `north' and 
 $\delta t\sim20\,$yr earlier in (the later of the two plates)
the `south'.  At that time, the lens and source
were separated by $\mu\delta t$.  If the lens is sufficiently bright, it will appear as a blob on the photographic plates and will therefore
 ``blot out'' the source at the epoch of the plate, and so the source 
will not appear in 
USNO-A2.0.  The exact blot-out radius depends on the magnitude of both the
lens and the source (fainter stars will get blotted-out farther from the 
lens).  However, the great majority of sources are relatively
faint ($V\sim 17$).  For simplicity, we therefore identify this radius as 
a function of lens magnitude, $\theta(V)$,
the point where 50\% of $V\sim 17$ stars are lost.  We find for 
$V=2, 5, 8, 11, 15$ that $\theta(V)=350, 80, 21, 11, 4$ arcseconds 
respectively.  Thus, for example, for a $V=8$
lens (i.e., $\theta=21''$), the minimum proper motion it is required to have to allow an event
to be detected is $\mu_{\rm min}=\theta(V)/\delta t=350\,\masyr$ in the
'north' or $1000\,\masyr$ in the `south'.

\section{Error Triage}

	The basic requirement for constructing a list of astrometric 
microlensing events
is to rank order the events by the amount of telescope time (here specifically
{\it SIM} time) needed to make a mass measurement of a specified precision.
At a later stage, one might decide to eliminate events with short observation
times because of some difficulty in carrying out the observations, and one
might choose to skip down the list to include an event with a long observation
time because the lens in question is exceptionally interesting.  However,
in this paper we will be concerned primarily with the fundamental requirement
of rank ordering the events.

	The observation time needed for a 1\% mass measurement
is given by (Gould 2000)
\begin{equation}
\tau = T_0 \alpha_0^2\biggl({r\beta c^2\over 4 G M}\biggr)^{2}\,
10^{0.4(V_s-17)}\,\gamma\biggl({\mu t_0
\over \beta},{\mu \Delta t\over \beta}
\biggr),
\label{eqn:taudef}
\end{equation}
where $r$ is the distance to the lens, $\beta$ is the impact parameter
of the event (the projected angular separation at the time $t_0$ of closest
approach), $M$ is the mass of the lens, $V_s$ is the apparent magnitude
of the source, $\mu$ is the relative lens-source proper motion, 
$\Delta t(=5\,$yr) is the duration of the experiment, $\gamma$ is a known
function which is discussed in detail by Gould (2000), 
$T_0=27\,{\rm hours}$, and $\alpha_0=100\,\mu$as.  

	In order to estimate $\tau$, one must first measure or estimate 
$r,\beta,M,V_s,\mu,$ and $t_0$.  Of course, there will be errors in all of
these quantities, and these will in turn generate errors in $\tau$.  In most
cases, these errors can be reduced by making additional observations or
carrying out additional investigations of various types.  However, these
refinements often require substantial legwork.  Therefore, one should first
decide what is an acceptable level of error in $\tau$ and what are the main contributors to it.

	The list of events will be constructed in three stages.  Stage 1 is
an automated search of a pair of star catalogs (sources and lenses) for events
with estimated observation times $\tau\leq\tau_{\rm max,1}$.  Stage 2 is
a simple (but potentially very time consuming) check of this list to eliminate
spurious candidates.  In stage 3, additional observations are made of the
remaining candidates.  The estimate of $\tau$ is refined and the final list
is constructed with a more restrictive maximum observation time
$\tau\leq\tau_{\rm max,3}$, and $\tau_{\rm max,3} < \tau_{\rm max,1}$.

	What level of errors are acceptable at stage 1 and stage 3?  At the
outset it should be emphasized that errors in the estimate of $\tau$ do
not cause errors in the final mass measurement by {\it SIM}.  The cost of errors
in stage 3 is that the {\it SIM} observations will be too short (causing larger
than desired statistical errors in the mass measurement) or too long
(wasting valuable {\it SIM} time pushing down the mass measurement errors
below what is actually desired). Hence, a factor of two error is acceptable.
That is, if the {\it SIM} time were underestimated by a factor of 2, then the
mass-measurement error would be 1.4\% instead of 1\%.  This would be a bit
worse than desired but on the other hand there would be a saving of {\it SIM} time
that could be applied to other stars.  If the {\it SIM} times were overestimated
by a factor of 2, then one would waste some {\it SIM} time on the event, but
one would reduce the error to 0.7\% which is not completely without value.
On the other hand, factor of 10 errors are not acceptable.  Either one would
waste a huge amount of {\it SIM} time, or one would obtain a mass measurement with
an error much larger than desired.  As a corollary, errors that are small
compared to a factor of 2 can be ignored at any stage.

	Much larger errors can be tolerated at stage 1 than stage 3.  For
example, if the stage-1 estimates could be in error by a factor of 10, then
one must set $\tau_{\rm max,1} = 10\tau_{\rm max,3}$ to avoid losing viable
candidates.  The cost is that the candidate list is increased by a factor
$(\tau_{\rm max,1}/\tau_{\rm max,3})^{1/2}\sim 3$ (Gould 2000), and one
must then sift through this larger list in stages 2 and 3.  Clearly, however,
this work load can become prohibitive for sufficiently large errors.

	We now show that of all the input parameters, only the distance $r$
and the impact parameter $\beta$ can induce sufficiently large uncertainties
in $\tau$ to warrant special attention.  We examine the various parameters
in turn.  

	If the lens is taken from the Hipparcos catalog, it will have
a trigonometric parallax.  In virtually all cases of interest, 
the lens will be close enough ($r\la 200\,$pc) that the distance error
will be less than 20\%, which is quite adequate for present purposes.  If
the lens does not have a trigonometric parallax, its distance must be
estimated from its measured flux (in say $V$ band) $F_V$ together with an
estimate of its intrinsic luminosity, $L_V$:
\begin{equation}
\tau\propto r^2 = {L_V\over 4\pi F_V}.
\label{eqn:rlum}
\end{equation}
Equation (\ref{eqn:rlum}) makes it appear as though the uncertainty in $\tau$
will be enormous.  For example, a star with a measured color $V-I=1$ could
plausibly be a clump giant with $M_V=1$, a main-sequence star with
$M_V=6$, a subdwarf with $M_V=8$, or a white dwarf with $M_V=14$.  This covers
a range of $1.6\times 10^5$ in luminosity and implies an uncertainty in $\tau$
of the same magnitude.  Nevertheless, we will show in \S\ 3.1
that with good two-band photometry, $r$ can be determined with $\sim 30\%$ accuracy 
which implies an error in $\tau$ of less than a factor of 2.  
Stars in the ACT catalog have good (Tycho) photometry.  For stars
in NLTT only photographic photometry is generally available.  We will show in 
\S\ 3.1 that for NLTT the $1\,\sigma$ errors in $L_V$ (and so $\tau$)
are a factor of 3.

	As discussed in \S\ 3.1, the first step in estimating the
distance to the lens is to determine its luminosity class (e.g., white dwarf,
subdwarf, main-sequence, or giant star).  If this is properly determined,
then the lens mass can be estimated quite accurately from the color.  For the
cases where the luminosity class is not correctly determined, the error
induced in the distance is much greater than the error induced in the mass.
Thus, in either case, the error in the mass can be ignored. 

	The geometry of the event ($\mu$, $\beta$, and $t_0$) is determined
from the astrometry.  These quantities affect the estimate of $\tau$ through 
the $\beta^2$ factor and the $\gamma$ factor in equation (\ref{eqn:taudef}).
We focus first on the $\beta^2$ factor.  As discussed in \S\ 2, the 
relative source-lens position error (and hence the error in $\beta$) is about
260 mas for lenses in Hipparcos and ACT and
about $1.\hskip-2pt ''2$ for NLTT.  In \S\ 4, we discuss how these
errors are incorporated into the search procedure.

	According to equation (\ref{eqn:taudef}) the 0.25 mag error in 
the source magnitude from USNO-A2.0 induces a 25\% error in $\tau$.  We ignore this.

	Finally, since the launch date of {\it SIM} is not fixed, we do not
attempt to calculate $\gamma$ based on the time of closest approach $t_0$
relative to the midpoint of the mission, 
$\gamma(\mu t_0/\beta,\mu\Delta t/\beta)$.  Rather, we calculate $\gamma$
for the optimal possible launch date for the given event when the midpoint of 
the mission
coincides with the time of closest approach, i.e., $t_0=0$.  That is, we use
$\gamma(0,\mu\Delta t/\beta)$.  Some representative values are
$\gamma(0,x)=10$ for $x\geq 4$,  $\gamma(0,2) = 19$, and $\gamma(0,1)=99$.
When the launch date is fixed and the time of minimum separation is better determined in the case of NLTT events, one can substitute the correct first argument
in place of 0.  In some cases, $\gamma$ may rise significantly but in others
(particularly when $\mu\Delta t\gg 4 \beta$) it will hardly be affected.
In any event, because we are suppressing consideration of the first argument,
any uncertainty in $t_0$ does not enter our calculation.

\subsection{Lens Distances}

	Here we describe our method for estimating the distances to the
lenses and evaluate the accuracy of these estimates.  Our method has three
distinct steps.  First, we assign a luminosity class to each star based on
its position in a reduced proper-motion diagram.  Second, we assign a $V$ band
luminosity $L_V$ (equivalently $M_V$)
to each star based on its luminosity class and color.  Third,
we combine the $L_V$ with the measured flux from the star $F_V$ 
(equivalently $V$) to obtain
a distance.  We apply this method to both the ACT and NLTT catalogs.  However,
to calibrate and describe the method, we first apply it to Hipparcos stars
with parallax errors smaller than 20\%.  After the method is calibrated, we
use it to ``predict'' the distances to these stars and then compare the
results to the measured Hipparcos parallaxes.

	Figure \ref{fig:one} is a reduced proper motion diagram of Hipparcos
stars with parallax errors smaller than 20\% (dots) and NLTT stars not present in Hipparcos (crosses).  (Please note that throughout 
this paper we will use $V$ magnitudes in Tycho system, and $B-V$ colors in 
Johnson system. To get Johnson $V$ magnitude, use the transformation (ESA, 1997): 
$V_{\rm J}=V-0.090(B-V)$. Consequently $M_V$ is in Tycho system as well.) 
If all stars had identical
transverse speeds $v_*$, then this diagram would look exactly like a
color-magnitude diagram (CMD), but with the vertical axis shifted by
$5\log (v_*/47.4\,\kms)$.  This means that disk stars (i.e., white
dwarfs, main-sequence stars, and giants) which have typical
$v_*\sim 30\,\kms$ are shifted upward by 1 mag, while halo stars
(i.e., subdwarfs) which have typical $v_*\sim 240\,\kms$
are shifted downward by 3.5 mag.  That is, the $\sim 2$ mag separation
between the main sequence and the subdwarfs in a ``normal'' CMD is here
augmented to $\sim 6.5$ mag. (But note that in a proper motion selected sample,
 like that from NLTT, there is a bias which makes the mean transverse speed of
 disk stars several times greater than their true speed $v_*$.)  We separate 
luminosity classes according to the
bold lines shown in the diagram.  Once these classes are chosen, we use 
color-magnitude relations for each class (white dwarfs, subdwarfs, 
main-sequence stars, and giants) to determine the absolute magnitude, and therefore the distance. Inevitably, stars of different class have some overlap in the reduced proper motion diagram, especially in the red end of the diagram. In those cases our classification is conservative, i.e. places a star in a class that will make it closer, and therefore producing smaller $\tau$.  

	Figure \ref{fig:two} compares the distances of Hipparcos stars derived
 from these luminosity estimates (together with the measured $V$ mags) to the 
true distances based on parallax.  For the typical lens distance moduli of 
less than 2.7 ($r=35\,$pc), the dispersion (excluding
outliers) is 0.53 mag. This is equivalent to a distance uncertainty of 28\%, 
and an error in the  estimate of $\tau$  of 63\%. For distance moduli greater than 2.7
, the dispersion is larger, but this is dominated by giants which are of 
little practical interest in the present search.

	We directly apply this technique to the ACT catalog for which there
is generally excellent photometry from Tycho.

	For NLTT, generally only photographic photometry is available.
Because of the large position errors in the NLTT catalog, we can search for
astrometric lensing events only if we can identify the NLTT star with
the corresponding object in USNO-A2.0.  Thus, in all cases we have photometry
from USNO-A2.0.  We convert from USNO-A2.0 mags
to Johnson color and Tycho $V$ using the relations:
\begin{equation}
B-V = 0.38+0.55(B_{\rm ph}-R_{\rm ph}),
\qquad V=R_{\rm ph}+0.23+0.32(B_{\rm ph}-R_{\rm ph}) ,
\label{eqn:usnojohnson1}
\end{equation}
The transformations were
derived by comparing  USNO-A2.0 photographic magnitudes to ground-based Johnson magnitudes of some sixty faint M dwarfs and white dwarfs. Here, $B_{\rm ph}$ and $R_{\rm ph}$ are blue and 
red photographic magnitudes, respectively. The 
scatter in the
predicted versus actual $B-V$ is $\sim 0.25$ mag.  Since the slope of the
main sequence is $\Delta V/\Delta(B-V)\sim 5$, the error in distance modulus of
NLTT stars
is $\sim 1.2\,$mag or about a factor of 1.7 in distance.  This corresponds
to a factor 3 error ($1\,\sigma$) in {\it SIM} time $\tau$.

	In the case of ACT and NLTT, the distance is also used to find the luminosity, that in turn (using mass-luminosity relations) determines the masses of main sequence stars and subdwarfs. For giants and white dwarfs we adopt masses of $1 M_{\odot}$ and  $0.6 M_{\odot}$ respectively.

	With Hipparcos, the mass is found directly because their distances,
 and therefore luminosities
 are known from trigonometric parallax. Thus, we only need to correctly 
determine the luminosity type based on luminosity and color (CMD). This is 
fairly straightforward for giants and white dwarfs, but can be ambiguous for 
main sequence stars vs. subdwarfs since they occupy not too different regions 
of CMD. We differentiate them by their transverse velocities, calling stars with $v>85\,\kms$ subdwarfs. Exact classification is only possible with additional information, such as a spectrum. In any case, as noted previously, the mass determination is not critical for the estimate of $\tau$. 

\subsection{Impact Parameter}  

	At first sight, the uncertainty in the impact parameter
(260 mas for Hipparcos and ACT, $1.\hskip-2pt ''2$ for NLTT) would appear
to wreak havoc with the estimate of $\tau$.  For example, any source 
whose calculated impact parameter with respect to an NLTT lens is less than
$2''$ might actually pass within 50 mas or even closer, thus reducing
its {\it SIM} time by a factor of 1600 or more.  In essence, one would seem to
be forced to do follow-up observations of all encounters in this catalog having apparent
impact parameters $\beta<2''$ in order to find the small
subset with very close encounters.  In fact, the situation is
not quite so severe.

	The size of the stop for {\it SIM} has not yet been fixed, but is likely to 
be about 300 mas.  This is about the size of the envelope of the {\it SIM} fringe
pattern (set by the 25 cm size of the mirrors).  Hence, if the lens is as
bright as the source then it would be difficult to obtain reliable 
astrometry while the source is within 300 mas of the lens.  Typically, the
lens will be much brighter than the source so the problem will be even more
severe.  For events with $\beta<300\,$mas, observations can be carried out
during most of the event, but must be suspended during the period of
closest approach.  The precision of the mass measurement will then be 
approximately the same as for an event with $\beta=300\,$mas.  That is,
there is an effective minimum impact parameter, $\beta_{\rm min}=300\,$mas.

	We account for the errors in distance and impact parameters as follows.
We aim for a final catalog with $\tau_{\rm max,3}=100\,$hrs.  For the
Hipparcos and ACT lenses, we accept the lens distances and impact parameters
at face value, but set $\tau_{\rm max,1}=300\,$hrs to allow for errors,
primarily overestimation of the impact parameter.  For NLTT we set
$\tau_{\rm max,1}=1000\,$hrs to allow for $1\,\sigma$ photometry errors.
In addition, we calculate the $\tau^*$ (the best-case $\tau$) by  reducing $\beta$ so that
\begin{equation}
\beta^*\rightarrow {\rm max}(\beta - 1.\hskip-2pt ''8,0'')
\label{eqn:betaadj}
\end{equation}
We always use the reduced $\beta^*$ to calculate the corresponding $\gamma$ factor that enters $\tau^*$, but in cases when $\beta^*<300\,$ mas, we use $\beta^*=300\,$ mas for the value of the impact parameter, because of the discussed aperture stop.
Finally, we allow all events where the lens is fainter than the source
and the nominal impact parameter is $\beta<1.\hskip-2pt ''8$ on the off
chance that the true impact parameter is very small. It is this best-case $\tau^*$ on which we impose the 1000 hr limit.
These three adjustments to the NLTT-based catalog mean that it will contain
a large number of spurious candidates.  These must be eliminated by follow-up
observations to obtain better photometry (which in some cases is available from the literature) and astrometry.

\section{Searching for the Candidate Events}

	Although the basic strategy for searching for the candidate events is
 the same for all three catalogs (Hipparcos, ACT and NLTT) there are some 
specific details that apply to each of them. Also, there were certain 
problems associated with the raw lists of events produced by these catalogs. That 
is, each catalog's initial list had its own set of `events' that turned out not to be real.

	The catalog of sources, USNO-A2.0, is written on 11 CD-ROMs, and the 
sky is divided into 24 zones each corresponding to $7.\hskip-2pt^\circ 5$ in 
declination. Each zone is written as one file. Our search program processes 
one zone at a time, checking every lens star that lies within that zone.

	First, the initial position of the lens in J2000.0 coordinates is 
needed. In the case of Hipparcos and ACT this is straightforward as they both 
list coordinates in the ICRS J2000.0 system, the one used by USNO-A2.0. One only 
needs to apply proper motion in order to change the epoch of the coordinates 
from 1991.25 and 2000.0 (Hipparcos and ACT, respectively) to that of the search period (2005-2015). In the case of 
NLTT, the procedure is much more involved. First, as explained in \S\ 2, we 
need to identify NLTT stars in USNO-A2.0 in order to get more accurate  
positions. We must therefore find a matching USNO-A2.0 star 
close to the position where NLTT was at the epoch of the {\it specific} plate 
that was scanned to produce entries in USNO-A2.0. This is essential because 
the span of plate epochs is quite wide, and the NLTT stars, having large 
proper motions, change their positions quickly. Therefore, in the first pass 
we look for anything close to where the NLTT star was at the mean epoch of 
entire POSS I (or SERC/ESO in the `south'). Since each USNO-A2.0 entry has a 
record of the plate from which it was scanned, we can determine the epoch from the table of plate epochs. With the exact 
epoch we know where precisely to look for an NLTT star. We do that by checking a $1.\hskip-2pt'5 
\times 1'$ error box ($\Delta_{\alpha}\times\Delta_{\delta}$), that accounts 
even for the worst initial 
NLTT positions. We accept as the best match a USNO-A2.0 star that is closest 
to the predicted position and has similar magnitude and color (in cases when 
NLTT lacks color information, only magnitude is used). In the `north', a
 match is found in  some 90\% of cases (90\% of which are within $10''$ of the expected 
position). In the great majority of cases when there is no match, the NLTT star was too faint
 to pass the detection limit of USNO-A2.0, or it was too bright, and therefore
 USNO-A2.0 had an entry with saturated photometry. However, the bright NLTT stars are almost always
recovered in Hipparcos and/or ACT. We made a special effort {\it not} to search NLTT stars that were included in Hipparcos or ACT: as discussed in \S\ 2, the NLTT data are of much lower quality and would generate many spurious events that are eliminated by the better Hipparcos and ACT data. We screen for these duplicates by
 looking for Hipparcos stars around NLTT positions that have similar proper motions ($\Delta \mu_\alpha,\, \Delta \mu_\delta < 40 \masyr$), and not too 
different magnitudes. We find 6233 matches, i.e., most of the Hipparcos stars 
with $\mu>200\,\masyr$. These matches are then flagged and skipped when 
identifying NLTT stars in USNO-A2.0. Also, if the match in USNO-A2.0 is 
associated with an ACT star, such NLTT star is also skipped. Occasionally, no match for an NLTT star is found because the 
input position was completely wrong [most likely a typo, since a 
machine-readable NLTT was produced by Optical Character Recognition (OCR)]. Identification efficiency is much 
worse in the `south' (SERC/ESO) for reasons discussed in \S\ 2. Only 20\%
 of NLTT stars are found within $10''$ of the expected position. 

	Next, the basic search strategy for events is to produce a box, the diagonal of 
which represents the lens's proper motion from 2005 to 2015, the time span during  
which an event should take place. The size of the box is further increased by 
5 years worth of proper motion (i.e, the largest possible impact parameter) 
to allow for events that take place near the starting and final years. We then find 
all the stars in USNO-A2.0 that are located within this box. A moving star, 
i.e. the lens, will pass by these stars, but not every encounter will produce 
a microlensing event. As discussed in \S\ 2 and \S\ 3 this depends on the 
physical parameters of the lens and on the brightness of the source star. Therefore, for each encounter we calculate the required {\it SIM} time and keep only events with  $\tau<\tau_{\rm max,1}$. In the case of NLTT, we use the reduced impact parameter $\beta^*$, as described in \S\ 3.2, to find $\tau^*$.

	Additionally, when searching ACT we discard encounters with stars that 
were labeled in USNO-A2.0 as being associated with ACT, in order to avoid 
finding encounters of an ACT star with `itself'. It might not sound logical to
 find an ACT star approaching its USNO-A2.0 entry in the {\it future}, but this 
happens with some slowly moving ACT stars because the astrometry of bright 
USNO-A2.0 stars is poor. A similar problem is present with bright Hipparcos 
stars, for 
which USNO-A2.0 sometimes contains multiple spurious entries. We discard 
these based on 
brightness and proximity of the Hipparcos star to the USNO-A2.0 entry at the epoch of the 
plate. Despite these automated rejection criteria, some `events' that are 
nothing other than the lens and its entry in USN0-A2.0, make their way into a 
final list. This most often happens because bright stars, having bad 
astrometry in USNO-A2.0, produce multiple entries if located in 
overlapping regions of the plates. These `events' are characterized by very short
 {\it SIM} observing times (because the `source' magnitude is bright). We 
check them by hand, by looking at the sky survey images themselves and making sure that there is only one star present.

	Once an event satisfying all criteria is found, the output list containing all the 
information about the lens, the source, and the geometry of an event is produced. We 
present these results in \S\ 5. However, the computer generated list is still 
far from containing only genuine events. One source of spurious entries affecting searches with Hipparcos and ACT catalogs is 
discussed in the preceding paragraph. Another problem is that since stars in these two catalogs
 are bright, their images in sky surveys have conspicuous diffraction spikes. 
These spikes in turn produce spurious entries in USNO-A2.0. Thus, sometimes an
 encounter will be reported in cases when the source is just an artifact 
from a diffraction spike. When we checked all of the Hipparcos and ACT events
 by comparing the sky survey images with USNO-A2.0 generated star charts ({\tt
 http://ftp.nofs.navy.mil/data/}), we were able to identify such occurrences. 
Also, since the diffraction spikes run along right ascension and declination, 
it was always 
the stars that had their proper motion along these directions that turned out
 to produce spurious events. 

	When it comes to NLTT, the most serious problem is with the encounters 
in the `south', because the lens identification is often spurious. These are 
checked by calculating how much the lens has moved between the two plates. If 
that distance is less than the $2''$ error circle (see \S\ 2) the chances are 
greater that the lens identification, and therefore the event, are real. Since
 there are not many of them, we check the `south' NLTT events by hand. Finally, since the NLTT position is sometimes completely off, it could lead to the wrong USNO star be identified as a match for NLTT star. Such a misidentified star might even produce an `event'. Since we do not check entries in NLTT list by hand, a possibility exists that some entries might not be real.

	As previously discussed, we try to eliminate doing NLTT stars that are present in either the Hipparcos or ACT catalogs. However, some survive our automated procedures. Therefore we check all NLTT events up to the Hipparcos/ACT detection limit and eliminate repetitions by hand. Thus, the NLTT list should contain only stars not present in the other two catalogs.

\section{Events}

	The events produced by stars in the Hipparcos and ACT catalogs are presented in Tables 1, 2 and 3. Tables 1 and 2 list the properties of the lens stars, while Table 3 lists those of the  source stars and of the events themselves. Details about specific columns are given in the table notes. The events are ordered by the required {\it SIM} time. There are 32 events taking place between years 2005 and 2015. Eight of them are found using both the Hipparcos and ACT catalogs (in which case the results presented are from the Hipparcos catalog), as indicated by the last column in Table 3. There was only one event (associated with the star AC368588) that was found in ACT and not in Hipparcos. However, inspection of POSS I and POSS II plates lead us to conclude that its proper motion is much smaller than that reported in ACT, and that no event will be taking place. One would expect all events detected by Hipparcos to be found in ACT, but this not the case. This is because in many cases of high-proper motion stars, the proper motion was not listed in Tycho, and therefore it is not listed in ACT either. In other cases, ACT was missing photometry because it was not available in Tycho. 

	These 32 events are produced by 25 different stars. Therefore, seven entries in Tables 1 and 2 are repetitions, but we keep them in order to preserve compatibility with Table 3, i.e., the `Event \#'.There are some notable stars among the lenses, such as Proxima Centauri (the closest star), Barnard Star (the highest proper motion), and the bright binary 61 Cyg A/B. They, together with the only white dwarf in the list (GJ 440), undergo multiple events that will both enable a more precise mass measurement and provide a check on systematics. 

	We classify 10 stars as subdwarfs, although some of them might be main sequence stars, and vice versa. A convenient way of presenting the types of Hipparcos stars that will undergo microlensing is given in Figure \ref{fig:three}. Plotted is the classical CMD of Hipparcos catalog stars with distances known to better than 10\%. Superimposed as big dots are the Hipparcos/ACT stars that produce events listed in Tables 1-3. As we see, except for a single white dwarf, the rest of the stars are uniformly distributed  within the faint ($M_V>5$) portion of the main sequence, with subdwarfs located mostly below the densest concentration of stars. The absence of stars with $M_V<5$ is the result of blotting out, as discussed in \S\ 2.

	Although the table includes events up to $\tau_{\rm max}=300\,$ hrs, they are concentrated towards shorter times. For example, $1/3$ of events have $\tau<20\,$ hrs, and $1/2$ less than 70 hrs. In fact, when we investigate the number of events as a function of $\tau$ we see a behavior that is in line with the theoretical predictions of Gould (2000).

	As an example, in Figure \ref{fig:four} we show the $8'\times8'$ field surrounding 61 Cyg A/B as it appeared in 1951 (DSS 1/POSS I) and in 1991 (DSS 2/POSS II) (upper left and lower left panels, respectively). We can see that the pair has moved some $3.\hskip-2pt'5$ across the field. In a  $2'\times2'$ blow-up we show the region that the pair will transverse in the period 2005-2015 (from DSS 2/POSS II). The star chart (created from USNO-A2.0 data), corresponds to the  $2'\times2'$ field and has the lensed stars labeled with the number of the corresponding event from Tables 1-3.

	Additional features of the set of events found with Hipparcos and ACT will be discussed later in this section, together with the events from NLTT.

	Tables 4 and 5 contain data about the 146 events found in NLTT, ordered by their nominal {\it SIM} observing time $\tau$. Details about the columns are given in the table notes. These tables have many more entries than the Hipparcos/ACT tables partly because of the $\tau^*_{\rm max}=1000\,$ hr limit compared to $\tau_{\rm max}=300\,$ hrs for Hipparcos/ACT. In fact, there are just 34 events with $\tau<300\,$ hrs. That means that if we had perfect knowledge about the NLTT stars there would be approximately 34 events in such a `perfect' list with $\tau<300\,$ hrs, but those, of course, would not necessarily be the first 34 from our present list. However, it should be noted that out of 146 events only 8 (5\%) are detected in the SERC/ESO part of USNO-A2.0 which comprises 35\% of the sky. Again, the nominal {\it SIM} observing times are concentrated toward the lower values, and the trend of the number of events vs. $\tau$ basically agrees with Gould (2000) predictions.

	NLTT events are produced evenly by stars that we classify as white dwarfs, subdwarfs, and late-type main-sequence stars. Such representation  is not surprising having in mind that most intrinsically bright, fast-moving stars are also apparently bright and therefore already covered by Hipparcos and ACT, so the ones covered by NLTT represent a sample of relatively nearby, intrinsically faint stars. One should keep in mind that our classification is conservative as not to miss a possible candidate, in the direction that some of our white dwarfs are actually subdwarfs or main-sequence stars, and some subdwarfs are main-sequence stars. This issue can be resolved in the stage 3 of list refinement, when better photometry and astrometry is obtained, supplemented by what is known about these stars from previous studies.

	Finally, both the Hipparcos/ACT and NLTT events can be investigated in the $V-\mu$ plane. This allows us to see the characteristics of the catalogs and events combined. Figure \ref{fig:five} covers a wide range of visual magnitudes ($2<V<19.5$) exhibited by high proper motion stars. It shows a range of proper motions from $\mu=0.\hskip-2pt''1\,{\rm yr}^{-1}$ to that of Barnard's star. The two long-dashed vertical lines show the nominal limit of the Hipparcos catalog of survey stars ($V=8$), and the detection limit of Hipparcos non-survey stars, Tycho, and therefore of ACT ($V=12$). The horizontal long-dashed line is the lower limit of $\mu= 0.\hskip-2pt''18\,{\rm yr}^{-1}$ for the NLTT. The lenses found only in Hipparcos are designated with `$\times$', and those found in both Hipparcos and ACT look like asterisks. In order to present a more realistic relative number of NLTT lenses, we plot only those with nominal $\tau<300\,$ hrs (circles). As discussed in \S\ 2, the blotting out of images in USNO-A2.0 limits our ability to find events moving slower than a specific value for the given lens magnitude. We plot this function $\theta(V)$ as a short-dashed line. Because of different epochs of POSS I and SERC/ESO, these cutoffs will be different in the two parts of the sky. The lower line corresponds to `north' (POSS I). The region below these two lines is therefore excluded, and we can see that none of the lens stars is found there. The exclusion due to blot-out approximately follows the diagonal line corresponding to a star with $M_V=6$, $v=75 \kms$. This shows that our survey cannot find disk-star lenses with $M_V<6$, unless they are moving faster than average. Indeed, as shown in Figure \ref{fig:three}, we find no lenses with $M_V\la 5$. However, halo stars with $M_V=6$, $v=240 \kms$ (upper diagonal line), are comfortably away from this limit.

\section{Conclusion and Discussion}

	Gould (2000) stressed the necessity of finding astrometric microlensing candidates to be observed by {\it SIM}, as soon as possible, since the separation between the lens and the source is steadily getting closer, and it will become harder to produce a valid estimate of the likelihood of an event the longer we wait. With the currently available catalogs, we were able to produce a fairly reliable list of candidates from Hipparcos and ACT catalogs. However, obtaining a list of similar quality of NLTT candidates requires additional astrometric and photometric observations of the candidates in our list. A one-meter class telescope with a CCD is adequate for such a job, since NLTT stars are relatively bright. Also, since obtaining an accurate color is  much more critical than a precise magnitude, the required observations can be successfully carried out in partially photometric conditions. Measuring the current relative separation of lens-source pair should refine the estimate of impact parameter sufficiently well, and requires a somewhat larger telescope. We are currently planning to carry out these observations on 2.4 m MDM Observatory telescope. 

	Another issue is getting more candidates. This can only be assured with new catalogs of proper motions, having lower proper motion cutoffs and going to fainter magnitudes. The biggest such projects are USNO-B and GSC II which should list the proper motions of  basically all the stars in POSS I/SERC/ESO. Having a lower proper motion limit is particularly important in $V>12$ range, where the blotting of stellar images no longer presents a limitation (at least not in the northern hemisphere), and where NLTT goes only to $\mu=180\,\masyr$. USNO-B will also push the detection limit $\sim1\,$ mag fainter compared to NLTT. Since in USNO-B all the stars will have proper motions, the uncertainty of the source star's position will also be reduced. Also, the completeness of NLTT at the fainter magnitudes is not altogether clear. According to  I.\ Reid (1999, private communication) it is actually only about 50\% complete near its proper-motion and magnitude limits. We did our own check by comparing the number of entries having $\mu>200\,\masyr$ in magnitude bin $V$ with the number of entries with $\mu>250\,\masyr$ in magnitude bin $V-0.5$.  In a perfectly complete catalog, the ratio of these two numbers should be 2 in each bin. Details of this completeness test are laid out in Flynn et al (1999).We see a significant drop only at $V>19$, i.e. we find NLTT to be complete. However, in the south and close to the galactic plane, the incompleteness sets already at $V\sim 14$. With USNO-B and GSC II this matter will most probably be resolved.

{\bf Acknowledgements}: 
We thank I.\ N. Reid for valuable discussions. Most of the computer-readable catalogs were obtained from Astronomical Data Center at NASA GSFC. This research was supported in part by grant AST 97-27520 from the NSF 
and in part by grant  NAG5-3111 from NASA.

\newpage

\begin{figure}
\caption[junk]{\label{fig:one}
Reduced proper motion diagram for Hipparcos and NLTT stars. In the case of Hipparcos, only stars with $\sigma_\pi/\pi<20\%$ are plotted (dots). From NLTT, only stars that are not in Hipparcos are plotted (crosses). In both cases, to avoid clutter, every 10th star is plotted.
 For Hipparcos data, abscissa is Johnson $B-V$ color as usually determined from Tycho photometry
but sometimes from ground-based photometry, while for NLTT it is calibrated from photographic magnitudes as given in catalog, and then randomized to correctly show regions of different density.  Ordinate is apparent magnitude
(Tycho $V$) augmented by the five times the 
logarithm of the proper motion in units of $''{\rm yr}^{-1}$, also known as the reduced proper motion.  If all stars had the
same transverse speed, this figure would look like an ordinary CMD.  Solid lines indicate
the boundaries of our assignment of stars to one of four classes: red giants are in the upper right corner, white dwarfs in the lower part, and subdwarfs between white dwarfs and the main sequence. 
}
\end{figure}

\begin{figure}
\caption[junk]{\label{fig:two}
Distance-modulus errors versus distance modulus for the Hipparcos stars shown
 in
 Fig.\ \ref{fig:one}.  The distance-modulus of each star is estimated by 
first classifying it according to the bold-line divisions in Fig.\
\ref{fig:one} and then assigning it an absolute magnitude using color-magnitude relations appropriate for each class.  The distance-modulus error is
then the difference between this estimate and the value based on the 
measured trigonometric parallax.  For distance moduli less than 2.7 (35 pc), the 
typical errors are only $\sim 0.53$ mag. Errors are larger for more distant stars, but these are dominated by giants which are not relevant in the present study.
}
\end{figure}

\begin{figure}
\caption[junk]{\label{fig:three}
A color-magnitude diagram of Hipparcos stars with distances measured to better than 10\%. The event-producing stars (lenses) from the Hipparcos and ACT catalogs are superimposed as filled dots, some of which are labeled.
}
\end{figure}

\begin{figure}
\caption[junk]{\label{fig:four}
$8'\times8'$ fields around 61 Cyg A/B in 1951 (upper left panel) and 1991 (lower left panel). Shown magnified is a $2'\times2'$ region where events will take place during 2005-2015 period. The chart corresponds to  the $2'\times2'$ field with source stars labeled with the numbers corresponding to `Event \#' in Tables 1-3. Events 6 and 25 are produced by 61 Cyg B, and the other three by 61 Cyg A.
}
\end{figure}

\begin{figure}
\caption[junk]{\label{fig:five}
Apparent magnitude - proper motion ($V-\mu$) plane showing lenses found in Hipparcos ($\times$), or both in Hipparcos and ACT catalogs ($*$). NLTT events that have $\tau<300$\, hr are shown as circles. Vertical long-dashed lines are completeness limits for Hipparcos and ACT. Horizontal long-dashed line is NLTT proper-motion cutoff. Short dashed lines delineate regions excluded due to blot-out (lower line - POSS I, upper line - SERC/ESO). Solid lines represent an $M_V=6$ star at various distances if belonging to proper motion selected disk population (lower line), or halo population (upper line).
}
\end{figure}

\def\kms{{\rm km}\,{\rm s}^{-1}}
\def\bmu{\hbox{$\mu\hskip-0.1pt\mu$}}
\begin{deluxetable}{l r r r r r r r r r r} 
 \tablecaption{Hipparcos and ACT events - lens star properties (astrometry and photometry)}
\scriptsize
\tablewidth{0pt}
\tablenum{1}
 \tablehead{
   \colhead{Event}      &
   \colhead{HIP \#}      &
   \multicolumn{3}{c}{RA}  &
   \multicolumn{3}{c}{DEC}  &
   \colhead{$V$} &
   \colhead{$B$$-$$V$} &
   \colhead{Other name} \\
   \colhead{\#} &
   \colhead{} &
   \colhead{h} &
   \colhead{m} &
   \colhead{s} &
   \colhead{$\circ$} &
   \colhead{$'$} &
   \colhead{$''$} &
   \colhead{} &
   \colhead{}  \\
 }
\startdata

1 & 104214 & 21 & 6 & 50.8350 & +38 & 44 & 29.380 & {\it 5.20}\phn & 1.069 & 61 Cyg A\\
2 & 106122 & 21 & 29 & 46.4600 & +45 & 53 & 37.083 & 7.986 & 0.759 & HD 204814\\
3 & 57367 & 11 & 45 & 39.2635 & $-$64 & 50 & 26.427 & 11.867 & 0.196 & GJ 440\\
4 & 90959 & 18 & 33 & 17.8712 & +22 & 18 & 55.449 & 9.016 & 1.181 & V774 Her\\
5 & 86214 & 17 & 37 & 4.2404 & $-$44 & 19 & 0.968 & {\it 10.94}\phn & 1.655 & GJ 682\\
6 & 104217 & 21 & 6 & 52.1924 & +38 & 44 & 3.890 & 6.208 & 1.309 & 61 Cyg B\\
7 & 28445 & 6 & 0 & 21.3792 & +31 & 25 & 50.855 & 9.505 & 0.930 & HD 250047\\
8 & 73734 & 15 & 4 & 19.2795 & +60 & 23 & 2.956 & {\it 11.00}\phn & 1.500 & Ross 1051\\
9 & 85523 & 17 & 28 & 39.4569 & $-$46 & 53 & 34.986 & {\it 9.38}\phn & 1.553 & GJ 674\\
10 & 104214 & 21 & 6 & 50.8350 & +38 & 44 & 29.380 & {\it 5.20}\phn & 1.069 & 61 Cyg A\\
11 & 70890 & 14 & 29 & 47.7474 & $-$62 & 40 & 52.867 & {\it 11.01}\phn & 1.807 & Proxima Cen\\
12 & 64965 & 13 & 18 & 57.0885 & $-$3 & 4 & 16.904 & {\it 10.84}\phn & 1.009 & Ross 484\\
13 & 57367 & 11 & 45 & 39.2635 & $-$64 & 50 & 26.427 & 11.867 & 0.196 & GJ 440\\
14 & 87937 & 17 & 57 & 48.9655 & +4 & 40 & 5.837 & {\it 9.54}\phn & 1.570 & Barnard's Star\\
15 & 74234 & 15 & 10 & 13.5770 & $-$16 & 27 & 15.521 & {\it 9.44}\phn & 0.850 & HD 134440\\
16 & 76074 & 15 & 32 & 13.8455 & $-$41 & 16 & 23.108 & {\it 9.31}\phn & 1.524 & GJ 588\\
17 & 98906 & 20 & 5 & 3.3563 & +54 & 26 & 11.144 & {\it 11.98}\phn & 1.524 & V1513 Cyg\\
18 & 61629 & 12 & 37 & 53.1966 & $-$52 & 0 & 5.580 & 10.767 & 1.470 & GJ 479\\
19 & 33582 & 6 & 58 & 38.3423 & $-$0 & 28 & 44.391 & 9.075 & 0.579 & HD 51754\\
20 & 114622 & 23 & 13 & 14.7435 & +57 & 10 & 3.498 & {\it 5.57}\phn & 1.000 & HD 219134\\
21 & 27207 & 5 & 46 & 1.5287 & +37 & 17 & 9.195 & 7.417 & 0.833 & HD 38230\\
22 & 74926 & 15 & 18 & 39.2706 & $-$18 & 37 & 32.607 & 10.643 & 1.214 & BD$-$18 4031\\
23 & 70890 & 14 & 29 & 47.7474 & $-$62 & 40 & 52.867 & {\it 11.01}\phn & 1.807 & Proxima Cen\\
24 & 70890 & 14 & 29 & 47.7474 & $-$62 & 40 & 52.867 & {\it 11.01}\phn & 1.807 & Proxima Cen\\
25 & 104217 & 21 & 6 & 52.1924 & +38 & 44 & 3.890 & 6.208 & 1.309 & 61 Cyg B\\
26 & 105090 & 21 & 17 & 17.7112 & $-$38 & 51 & 52.468 & {\it 6.69\phn} & 1.397 & AX Mic\\
27 & 102923 & 20 & 51 & 6.5386 & +7 & 1 & 40.380 & 10.014 & 0.900 & BD+06 4665\\
28 & 104214 & 21 & 6 & 50.8350 & +38 & 44 & 29.380 & {\it 5.20}\phn & 1.069 & 61 Cyg A\\
29 & 87937 & 17 & 57 & 48.9655 & +4 & 40 & 5.837 & {\it 9.54}\phn & 1.570 & Barnard's Star\\
30 & 79537 & 16 & 13 & 49.4874 & $-$57 & 34 & 1.492 & {\it 7.53}\phn & 0.815 & HD 145417\\
31 & 48336 & 9 & 51 & 8.9608 & $-$12 & 19 & 34.728 & 10.093 & 1.446 & SAO 155530\\
32 & 25878 & 5 & 31 & 26.9506 & $-$3 & 40 & 19.712 & 8.144 & 1.474 & HD 36395\\
\enddata
\tablenotetext{}{Rows are ordered by increasing $\tau$ (see Table 3). HIP \# is the Hipparcos catalog number. Right ascension and declination are taken from Hipparcos catalog. Equinox J2000, epoch 1991.25. Visual magnitude is in Tycho system if available, or Johnson (italics). Johnson colors are from Hipparcos catalog as well. Multiple entries arise from the fact that a single lens can produce multiple events.}
\normalsize
\end{deluxetable}

\def\kms{{\rm km}\,{\rm s}^{-1}}
\def\bmu{\hbox{$\mu\hskip-0.1pt\mu$}}
\begin{deluxetable}{l r r r r r r l l} 
 \tablecaption{Hipparcos and ACT events - lens star properties (distance, kinematic and physical properties)}
\scriptsize
\tablewidth{0pt}
\tablenum{2}
 \tablehead{
   \colhead{Event}      &
   \colhead{$\mu$} &
   \colhead{p.a.} &
   \colhead{$v_{\rm rad}$} &
   \colhead{$M_V$} &
   \colhead{$r$} &
   \colhead{$M$} &
   \colhead{Class} &
   \colhead{Sp} \\
   \colhead{\#} &
   \colhead{	 $\arcsec/{\rm\,yr}$} &
   \colhead{} &
   \colhead{$\kms$} &
   \colhead{} &
   \colhead{pc}  &
   \colhead{$M_\odot$}  \\
 }
\startdata

1 & 5.2807 & 
52 & $-$64.28 & 7.5 & 3.5 & 0.5 & SD & K5V\\
2 & 0.5531 &
 50 & $-$83.70 & 5.6 & 29.8 & 0.9 & MS & G8V\\
3 & 2.6876 &
 97 &  & 13.5 & 4.6 & 0.6 & WD & DC:\\
4 & 0.5052 & 
200 & 37.10 & 7.2 & 23.4 & 0.8 & MS & K4V\\
5 & 1.1765 & 
217 & $-$60.00 & 12.4 & 5.0 & 0.2 & MS & M5\\
6 & 5.1724 & 53
 & $-$63.48 & 8.5 & 3.5 & 0.4 & SD & K7V\\
7 & 0.3297 & 
155 &  & 6.2 & 46.1 & 0.9 & MS & K2\\
8 & 0.6786 & 
285 & & 9.8 & 17.6 & 0.5 & MS & M:\\
9 & 1.0501 &
 147 &  & 11.1 & 4.5 & 0.4 & MS & K5\\
10 & 5.2807 & 
52 & $-$64.28 & 7.5 & 3.5 & 0.5 & SD & K5V\\
11 & 3.8530 
& 281 &  & 15.4 & 1.3 & 0.1 & MS & M5Ve\\
12 & 0.6517 & 
258 & 126.00 & 8.1 & 35.9 & 0.5 & SD & K5\\
13 & 2.6876 
& 97 &  & 13.5 & 4.6 & 0.6 & WD & DC:\\
14 & 10.3577 & 
356 & $-$106.76 & 13.2 & 1.8 & 0.1 & SD & sdM4\\
15 & 3.6815 &
 196 & 308.08 & 7.1 & 29.7 & 0.6 & SD & K0V:\\
16 & 1.5636 &
 229 &  & 10.4 & 5.9 & 0.5 & MS & M0\\
17 & 1.4724 & 
232 & 0.01 & 11.0 & 15.8 & 0.2 & SD & M3\\
18 & 1.0347 & 
272 &  & 10.8 & 9.7 & 0.4 & MS & M3\\
19 & 0.6930 & 
151 & $-$80.59 & 4.9 & 68.4 & 0.8 & SD & G0\\
20 & 2.0952 &
 82 & $-$17.79 & 6.5 & 6.5 & 0.8 & MS & K3Vvar\\
21 & 0.7050 & 
136 & $-$30.90 & 5.9 & 20.6 & 0.9 & MS & K0V\\
22 & 0.5713 
& 128 &  & 8.5 & 26.2 & 0.6 & MS &\\
23 & 3.8530 
& 281 &  & 15.4 & 1.3 & 0.1 & MS & M5Ve\\
24 & 3.8530 
& 281 &  & 15.4 & 1.3 & 0.1 & MS & M5Ve\\
25 & 5.1724 & 
53 & $-$63.48 & 8.5 & 3.5 & 0.4 & SD & K7V\\
26 & 3.4549 
& 251 & 23.01 & 8.7 & 3.9 & 0.6 & MS & M1/M2V\\
27 & 0.4333 & 
147 &  & 6.6 & 48.3 & 0.6 & SD & K3\\
28 & 5.2807 &
 52 & $-$64.28 & 7.5 & 3.5 & 0.5 & SD & K5V\\
29 & 10.3577 & 
356 & $-$106.76 & 13.2 & 1.8 & 0.1 & SD & sdM4\\
30 & 1.6491 & 
211 & 10.01 & 6.8 & 13.7 & 0.6 & SD & K0V\\
31 & 1.8487 & 
142 & 61.01 & 9.4 & 13.7 & 0.4 & SD & M0\\
32 & 2.2277 & 
160 & 10.61 & 9.4 & 5.7 & 0.6 & MS & M1V\\
\enddata
\tablenotetext{}{ Proper motions are given as intensity and position angle (from Hipparcos). Radial velocities are taken from SIMBAD. Absolute magnitude is in the same system as the corresponding visual magnitude. Distances come from Hipparcos trigonometric parallaxes. For mass estimate and class determination see \S\ 3.1 (MS - main sequence star, SD - subdwarf, WD - white dwarf). Spectral class is taken from Hipparcos catalog.}
\normalsize
\end{deluxetable}

\def\kms{{\rm km}\,{\rm s}^{-1}}
\begin{deluxetable}{l r r r r r r r r r r r r r} 
 \tablecaption{Hipparcos and ACT events - source star and event properties}
\scriptsize
\tablewidth{0pt}
\tablenum{3}
 \tablehead{
   \colhead{Event}      &
   \multicolumn{3}{c}{RA}  &
   \multicolumn{3}{c}{DEC}  &
   \colhead{$V$} &
   \colhead{$B$$-$$V$} &
   \colhead{$\tau$} &
   \colhead{$d_{2000}$} &
   \colhead{$t_0$} &
   \colhead{$\beta$} \\
   \colhead{\#} &
   \colhead{h} &
   \colhead{m} &
   \colhead{s} &
   \colhead{$\circ$} &
   \colhead{$'$} &
   \colhead{$''$} &
   \colhead{} &
   \colhead{} &
   \colhead{hr} &
   \colhead{$''$} &
   \colhead{yr} &
   \colhead{mas} \\
 }
\startdata

1 & 21 & 6 & 58.229 & +38 & 45 & 41.14 & 10.7 &  & 0.1 & 66.2 & 2012.5 & 3064 & H\\
2 & 21 & 29 & 47.366 & +45 & 53 & 45.37 & 16.6 & 1.43 & 4.0 & 7.7 & 2013.9 & 366 & H/A\\
3 & 11 & 45 & 48.968 & $-$64 & 50 & 33.74 & 18.2 & 1.26 & 4.2 & 38.8 & 2014.4 & 738 & H\\
4 & 18 & 33 & 17.603 & +22 & 18 & 46.65 & 16.4 & 0.22 & 5.0 & 5.2 & 2010.2 & 456 & H/A\\
5 & 17 & 37 & 2.520 & $-$44 & 19 & 21.96 & 13.7 & 1.37 & 5.4 & 17.7 & 2014.9 & 2022 & H\\
6 & 21 & 6 & 59.946 & +38 & 45 & 14.23 & 18.8 & 0.93 & 5.5 & 69.6 & 2013.5 & 636 & H\\
7 & 6 & 0 & 21.624 & +31 & 25 & 44.73 & 17.0 & 1.76 & 10.7 & 4.0 & 2012.2 & 262 & H/A\\
8 & 15 & 4 & 17.808 & +60 & 23 & 5.38 & 17.1 & 0.77 & 12.6 & 5.2 & 2007.7 & 520 & H\\
9 & 17 & 28 & 40.850 & $-$46 & 53 & 50.64 & 13.8 & 1.92 & 12.9 & 12.2 & 2011.2 & 3401 & H\\
10 & 21 & 6 & 56.938 & +38 & 45 & 20.65 & 16.3 & 1.37 & 13.2 & 41.8 & 2007.9 & 3686 & H\\
11 & 14 & 29 & 39.583 & $-$62 & 40 & 42.81 & 17.2 & 1.70 & 13.3 & 23.4 & 2006.1 & 1360 & H\\
12 & 13 & 18 & 56.199 & $-$3 & 4 & 20.84 & 13.9 & 0.77 & 21.8 & 8.3 & 2012.6 & 1136 & H\\
13 & 11 & 45 & 46.125 & $-$64 & 50 & 29.29 & 17.8 & 1.26 & 41.6 & 20.4 & 2007.5 & 2805 & H\\
14 & 17 & 57 & 47.972 & +4 & 43 & 0.77 & 18.8 & 0.49 & 58.0 & 83.3 & 2008.2 & 1286 & H\\
15 & 15 & 10 & 11.993 & $-$16 & 28 & 29.09 & 14.8 & 0.88 & 58.1 & 44.8 & 2012.2 & 1862 & H\\
16 & 15 & 32 & 11.805 & $-$41 & 16 & 47.52 & 16.5 & 1.37 & 67.2 & 20.0 & 2012.6 & 3261 & H\\
17 & 20 & 5 & 1.087 & +54 & 25 & 56.10 & 18.7 & 1.10 & 87.9 & 12.0 & 2008.2 & 243 & H\\
18 & 12 & 37 & 50.912 & $-$52 & 0 & 5.95 & 18.6 & 1.48 & 91.7 & 12.1 & 2011.6 & 974 & H/A\\
19 & 6 & 58 & 38.763 & $-$0 & 28 & 53.97 & 16.1 & 0.93 & 92.0 & 5.4 & 2007.8 & 863 & H/A\\
20 & 23 & 13 & 18.727 & +57 & 10 & 12.54 & 17.2 & 1.26 & 92.8 & 15.7 & 2007.2 & 4364 & H\\
21 & 5 & 46 & 2.356 & +37 & 17 & 0.80 & 17.8 & 1.21 & 113.5 & 6.9 & 2009.6 & 1349 & H/A\\
22 & 15 & 18 & 40.012 & $-$18 & 37 & 39.53 & 17.1 & 0.82 & 133.3 & 7.6 & 2013.2 & 1109 & H/A\\
23 & 14 & 29 & 38.124 & $-$62 & 40 & 35.97 & 17.9 & 0.99 & 147.4 & 34.7 & 2009.0 & 3341 & H\\
24 & 14 & 29 & 36.081 & $-$62 & 40 & 40.49 & 17.6 & 1.10 & 157.5 & 47.6 & 2012.3 & 3885 & H\\
25 & 21 & 6 & 58.044 & +38 & 44 & 44.13 & 16.4 & 0.66 & 187.4 & 34.9 & 2006.5 & 9655 & H\\
26 & 21 & 17 & 13.381 & $-$38 & 51 & 59.53 & 16.0 & 1.92 & 192.1 & 22.3 & 2005.7 & 10139 & H\\
27 & 20 & 51 & 6.792 & +7 & 1 & 35.53 & 18.3 & 0.88 & 194.3 & 2.4 & 2005.4 & 464 & H/A\\
28 & 21 & 6 & 58.833 & +38 & 45 & 32.30 & 17.4 & 1.70 & 221.6 & 66.8 & 2012.6 & 8266 & H\\
29 & 17 & 57 & 47.361 & +4 & 44 & 5.21 & 16.7 & 1.10 & 231.6 & 150.0 & 2014.5 & 5503 & H\\
30 & 16 & 13 & 47.751 & $-$57 & 34 & 27.26 & 18.7 & 1.76 & 242.9 & 14.9 & 2009.0 & 1414 & H\\
31 & 9 & 51 & 10.020 & $-$12 & 19 & 57.92 & 17.0 & 0.49 & 261.4 & 11.8 & 2006.3 & 2040 & H\\
32 & 5 & 31 & 27.623 & $-$3 & 40 & 56.32 & 19.1 & 0.71 & 291.2 & 18.6 & 2008.2 & 3056 & H\\
\enddata
\tablenotetext{}{Numeration follows the numbers in tables 1 and 2. Source star's right ascension and declination are from USNO-A2.0, at plate epoch, equinox J2000. Visual magnitude is in Tycho, and color in Johnson system, as calibrated from photographic magnitudes (see \S\ 3.1). Event is described by $\tau$ ({\it SIM} observing time), $d_{2000}$ lens-source separation in year 2000.0, $t_0$ time of closest approach and $\beta$, the minimum impact parameter. If $\beta<300\,$mas, $\tau$ is calculated using  $\beta=300\,$mas. The last column designates whether the event was detected only using the Hipparcos catalog (H), or in both the Hipparcos and the ACT catalogs (H/A).}
\normalsize
\end{deluxetable}

\def\kms{{\rm km}\,{\rm s}^{-1}}
\def\bmu{\hbox{$\mu\hskip-0.1pt\mu$}}
\begin{deluxetable}{l l r r r r r r r r r r r r r r r} 
 \tablecaption{NLTT events - lens star properties }
\tiny
\tablewidth{0pt}
\tablenum{4}
 \tablehead{
   \colhead{Event}      &
   \colhead{Name}      &
   \multicolumn{3}{c}{RA}  &
   \multicolumn{3}{c}{DEC}  &
   \colhead{Epoch} &
   \colhead{$V$} &
   \colhead{$B$$-$$V$} &
   \colhead{$\mu$} &
   \colhead{p.a.} &
   \colhead{$M_V$} &
   \colhead{$r$} &
   \colhead{$M$} &
   \colhead{Class} \\
   \colhead{\#} &
   \colhead{} &
   \colhead{h} &
   \colhead{m} &
   \colhead{s} &
   \colhead{$\circ$} &
   \colhead{$'$} &
   \colhead{$''$} &
   \colhead{}  &
   \colhead{} &
   \colhead{} &
   \colhead{         $\arcsec/{\rm\,yr}$} &
   \colhead{} &
   \colhead{} &
   \colhead{pc}  &
   \colhead{$M_\odot$}  \\
}
\startdata
1 &  & 0 & 35 & 49.472 & +52 & 41 & 20.43 & 1954.8 & 12.3 & 1.75 & 0.789 & 102 & 14.9 & 3 & 0.1 & MS \\
2 & 452-  1 & 19 & 21 & 41.879 & +20 & 53 & 11.01 & 1953.6 & 13.6 & 1.97 & 1.751 & 213 & 18.8 & 1 & 0.1 & MS \\
3 &  & 23 & 6 & 20.173 & +65 & 3 & 40.05 & 1952.6 & 15.4 & 1.92 & 0.328 & 127 & 17.8 & 3 & 0.1 & MS \\
4 & 755- 18 & 20 & 27 & 30.371 & $-$13 & 17 & 36.29 & 1953.8 & 18.2 & 0.66 & 0.375 & 215 & 15.0 & 46 & 0.6 & WD \\
5 & 650-237 & 2 & 31 & 56.521 & $-$8 & 31 & 49.98 & 1953.9 & 16.0 & 1.54 & 0.302 & 164 & 11.0 & 104 & 0.2 & SD \\
6 & 634-  1 & 19 & 56 & 30.645 & $-$1 & 1 & 58.61 & 1951.6 & 13.6 & 0.49 & 0.790 & 211 & 14.5 & 7 & 0.6 & WD \\
7 & 275- 67 & 16 & 35 & 14.667 & +35 & 47 & 26.59 & 1954.5 & 13.8 & 1.98 & 0.221 & 236 & 18.8 & 1 & 0.1 & MS \\
8 & R 627 & 11 & 24 & 16.301 & +21 & 21 & 36.50 & 1955.2 & 14.3 & 0.76 & 1.050 & 270 & 15.1 & 7 & 0.6 & WD \\
9 & 543- 33 & 7 & 50 & 14.730 & +7 & 12 & 55.88 & 1956.0 & 16.7 & 0.99 & 1.778 & 173 & 15.3 & 19 & 0.6 & WD \\
10 &  & 23 & 18 & 6.829 & +49 & 28 & 28.63 & 1954.6 & 13.2 & 1.15 & 0.320 & 177 & 8.0 & 107 & 0.5 & SD \\
11 & R 619 & 8 & 11 & 54.025 & +8 & 50 & 31.39 & 1951.2 & 13.0 & 1.37 & 5.211 & 167 & 15.9 & 3 & 0.6 & WD \\
12 & 388- 57* & 17 & 36 & 11.388 & +23 & 48 & 32.58 & 1951.5 & 19.4 & 1.21 & 0.184 & 196 & 15.6 & 59 & 0.6 & WD \\
13 & R 248 & 23 & 41 & 54.568 & +44 & 11 & 54.44 & 1952.6 & 12.8 & 2.08 & 1.617 & 177 & 20.8 & 0 & 0.1 & MS \\
14 & 707-  8 & 1 & 9 & 2.851 & $-$10 & 42 & 12.41 & 1951.9 & 16.5 & 0.44 & 0.198 & 98 & 14.3 & 27 & 0.6 & WD \\
15 & 816- 34 & 21 & 0 & 36.541 & $-$18 & 16 & 49.96 & 1982.5 & 17.1 & 0.71 & 0.198 & 207 & 15.0 & 26 & 0.6 & WD \\
16 & 385- 32 & 16 & 6 & 36.066 & +24 & 28 & 56.74 & 1950.4 & 18.4 & 1.43 & 0.316 & 191 & 16.1 & 28 & 0.6 & WD \\
17 &  & 5 & 50 & 22.925 & +17 & 19 & 40.47 & 1951.9 & 15.9 & 0.60 & 0.589 & 143 & 14.8 & 16 & 0.6 & WD \\
18 & 197-  4 & 2 & 25 & 40.398 & +42 & 27 & 9.20 & 1952.0 & 18.4 & 1.10 & 0.232 & 103 & 15.4 & 40 & 0.6 & WD \\
19 & 795- 43 & 12 & 38 & 42.281 & $-$19 & 21 & 39.48 & 1954.2 & 13.1 & 1.53 & 0.356 & 301 & 11.0 & 27 & 0.4 & MS \\
20 &  & 19 & 38 & 48.802 & +35 & 11 & 59.24 & 1952.5 & 15.0 & 1.48 & 0.786 & 359 & 10.0 & 101 & 0.3 & SD \\
21 & 329- 21 & 16 & 1 & 47.561 & +30 & 30 & 56.40 & 1950.4 & 19.1 & 1.21 & 0.217 & 151 & 15.6 & 51 & 0.6 & WD \\
22 & 689- 11 & 17 & 55 & 49.486 & $-$7 & 35 & 52.55 & 1954.5 & 12.4 & 1.31 & 0.253 & 234 & 8.2 & 70 & 0.7 & MS \\
23 & 101- 15* & 16 & 34 & 26.512 & +57 & 8 & 51.70 & 1955.3 & 13.2 & 1.59 & 1.620 & 316 & 11.9 & 18 & 0.2 & SD \\
24 & * & 1 & 47 & 57.166 & +60 & 7 & 37.37 & 1954.8 & 13.4 & 1.59 & 0.238 & 228 & 11.9 & 19 & 0.3 & MS \\
25 & W 1471 & 17 & 42 & 13.355 & $-$8 & 48 & 38.47 & 1954.5 & 13.5 & 1.64 & 0.965 & 240 & 12.9 & 13 & 0.2 & MS \\
26 & 332- 17 & 17 & 19 & 4.649 & +28 & 5 & 10.42 & 1950.5 & 16.8 & 1.76 & 0.237 & 239 & 14.9 & 24 & 0.1 & MS \\
27 & 497-  4 & 13 & 8 & 26.657 & +12 & 26 & 37.14 & 1955.4 & 14.3 & 1.70 & 0.288 & 268 & 13.9 & 12 & 0.1 & MS \\
28 & R 201 & 21 & 40 & 27.505 & +54 & 0 & 27.20 & 1955.9 & 14.9 & 1.59 & 0.414 & 76 & 11.9 & 39 & 0.3 & MS \\
29 & R  28 & 4 & 13 & 0.374 & +52 & 37 & 18.86 & 1954.8 & 13.7 & 1.65 & 0.910 & 203 & 12.9 & 14 & 0.2 & MS \\
30 & 921- 25 & 18 & 6 & 31.135 & $-$30 & 9 & 46.02 & 1977.5 & 16.1 & 0.16 & 0.252 & 158 & 12.7 & 48 & 0.6 & WD \\
31 & 627- 16 & 17 & 15 & 24.814 & +1 & 19 & 17.37 & 1954.6 & 15.6 & 1.59 & 0.369 & 287 & 11.9 & 53 & 0.2 & SD \\
32 & 29- 23 & 0 & 43 & 57.017 & +75 & 12 & 26.58 & 1954.7 & 18.3 & 0.82 & 0.302 & 104 & 15.2 & 41 & 0.6 & WD \\
33 & 785- 11 & 8 & 31 & 8.147 & $-$20 & 41 & 59.95 & 1982.0 & 17.1 & 1.75 & 0.251 & 132 & 14.9 & 28 & 0.1 & MS \\
34 & 575- 26 & 20 & 34 & 31.828 & +7 & 57 & 32.55 & 1951.6 & 15.0 & 1.75 & 0.374 & 82 & 14.9 & 11 & 0.1 & MS \\
35 & 722-  1 & 7 & 13 & 39.010 & $-$13 & 27 & 8.92 & 1958.9 & 14.7 & 1.21 & 1.277 & 153 & 15.6 & 7 & 0.6 & WD \\
36 &  & 5 & 10 & 28.686 & +31 & 17 & 40.41 & 1955.8 & 17.0 & 1.21 & 0.690 & 104 & 15.6 & 19 & 0.6 & WD \\
37 & 48-813 & 23 & 5 & 14.276 & +71 & 23 & 4.05 & 1952.6 & 19.4 & 0.60 & 0.277 & 56 & 14.8 & 82 & 0.6 & WD \\
38 & W 1084 & 20 & 43 & 14.497 & +55 & 19 & 31.92 & 1952.7 & 15.1 & 1.59 & 1.915 & 28 & 16.9 & 4 & 0.6 & WD \\
39 & 447- 63 & 17 & 7 & 15.733 & +19 & 25 & 51.92 & 1954.5 & 13.3 & 1.65 & 0.180 & 166 & 12.9 & 12 & 0.2 & MS \\
40 & 206- 11 & 7 & 11 & 9.967 & +43 & 30 & 24.24 & 1954.2 & 15.8 & 1.59 & 0.680 & 146 & 11.9 & 59 & 0.2 & SD \\
41 & 572-  1 & 19 & 22 & 4.534 & +7 & 2 & 51.55 & 1950.6 & 12.5 & 1.81 & 0.836 & 242 & 15.9 & 2 & 0.1 & MS \\
42 &  & 22 & 36 & 37.955 & +53 & 3 & 16.61 & 1953.8 & 17.2 & 0.05 & 0.260 & 226 & 11.7 & 127 & 0.6 & WD \\
43 & 727-  3 & 9 & 9 & 54.157 & $-$11 & 26 & 6.38 & 1954.2 & 14.7 & 1.65 & 0.483 & 118 & 12.9 & 23 & 0.2 & MS \\
44 &  & 3 & 43 & 49.491 & +63 & 40 & 30.53 & 1954.1 & 12.9 & 1.37 & 0.962 & 142 & 8.5 & 77 & 0.4 & SD \\
45 & Stein 2051B* & 4 & 31 & 1.299 & +59 & 0 & 32.22 & 1953.1 & 13.5 & 1.15 & 2.383 & 144 & 15.5 & 4 & 0.6 & WD \\
46 &  & 0 & 9 & 52.296 & +53 & 1 & 12.88 & 1953.0 & 13.5 & 2.08 & 0.240 & 85 & 20.8 & 0 & 0.1 & MS \\
47 & 106- 38 & 20 & 14 & 42.347 & +61 & 46 & 2.31 & 1952.7 & 16.9 & 1.43 & 0.698 & 26 & 16.1 & 14 & 0.6 & WD \\
48 &  & 20 & 34 & 2.957 & +64 & 19 & 16.39 & 1952.6 & 13.5 & 1.43 & 0.436 & 254 & 9.0 & 78 & 0.4 & SD \\
49 & 297- 12 & 2 & 11 & 57.902 & +32 & 21 & 56.99 & 1954.8 & 15.9 & 1.87 & 0.567 & 114 & 16.8 & 6 & 0.1 & MS \\
50 &  & 2 & 7 & 2.738 & +49 & 39 & 3.27 & 1953.9 & 13.0 & 1.15 & 0.498 & 150 & 8.0 & 98 & 0.5 & SD \\

\tablebreak
51 & 399-299 & 22 & 1 & 5.921 & +29 & 9 & 37.24 & 1951.7 & 16.2 & 0.82 & 0.598 & 94 & 15.2 & 16 & 0.6 & WD \\
52 & W 359* & 10 & 56 & 41.064 & +7 & 2 & 59.26 & 1953.3 & 13.9 & 2.14 & 4.696 & 234 & 21.7 & 0 & 0.1 & MS \\
53 &  & 1 & 4 & 2.104 & +59 & 38 & 5.31 & 1952.7 & 15.1 & 1.65 & 0.420 & 100 & 12.9 & 28 & 0.2 & MS \\
54 & T 9 & 17 & 18 & 46.499 & $-$29 & 46 & 5.89 & 1981.4 & 13.1 & 1.97 & 0.242 & 109 & 18.8 & 1 & 0.1 & MS \\
55 &  & 1 & 48 & 47.971 & +55 & 2 & 7.76 & 1954.8 & 14.4 & 1.26 & 0.280 & 96 & 8.3 & 165 & 0.4 & SD \\
56 & +15:4074B* & 20 & 11 & 14.683 & +16 & 10 & 48.99 & 1951.7 & 13.8 & 1.42 & 0.572 & 313 & 9.0 & 90 & 0.4 & SD \\
57 & 693- 14 & 19 & 38 & 31.486 & $-$2 & 51 & 12.09 & 1953.8 & 11.1 & 0.83 & 0.286 & 111 & 7.4 & 55 & 0.5 & SD \\
58 & 685- 55 & 16 & 34 & 41.539 & $-$9 & 1 & 44.30 & 1954.3 & 12.2 & 0.60 & 0.185 & 176 & 6.1 & 167 & 0.7 & SD \\
59 & 569- 98 & 18 & 2 & 32.457 & +5 & 45 & 2.87 & 1950.5 & 18.8 & 0.76 & 0.472 & 205 & 15.1 & 55 & 0.6 & WD \\
60 & R  19 & 2 & 19 & 0.361 & +35 & 21 & 39.42 & 1951.8 & 12.7 & 1.65 & 0.792 & 122 & 12.9 & 9 & 0.2 & MS \\
61 &  & 12 & 38 & 33.755 & +35 & 13 & 19.73 & 1950.4 & 14.9 & 1.53 & 0.267 & 223 & 11.0 & 63 & 0.4 & MS \\
62 &  & 0 & 28 & 51.956 & +50 & 22 & 27.91 & 1954.7 & 13.3 & 1.98 & 0.440 & 74 & 18.8 & 1 & 0.1 & MS \\
63 & 555-  5 & 12 & 21 & 49.935 & +6 & 44 & 7.77 & 1956.2 & 14.5 & 1.53 & 0.732 & 174 & 11.0 & 52 & 0.2 & SD \\
64 & +53:2911* & 22 & 32 & 49.386 & +53 & 47 & 35.92 & 1952.7 & 10.0 & 1.19 & 1.318 & 86 & 8.1 & 24 & 0.5 & SD \\
65 & 44- 47 & 17 & 37 & 24.422 & +71 & 4 & 16.12 & 1953.7 & 12.5 & 1.53 & 0.482 & 143 & 11.0 & 21 & 0.4 & MS \\
66 & 751-  1 & 19 & 3 & 17.340 & $-$13 & 33 & 51.30 & 1951.6 & 16.4 & 1.53 & 0.780 & 226 & 16.6 & 9 & 0.6 & WD \\
67 &  & 5 & 44 & 0.906 & +40 & 57 & 36.75 & 1953.0 & 15.8 & 0.33 & 1.229 & 147 & 13.8 & 25 & 0.6 & WD \\
68 &  & 21 & 35 & 19.123 & +46 & 33 & 41.32 & 1952.7 & 17.2 & 0.93 & 0.459 & 200 & 15.3 & 25 & 0.6 & WD \\
69 & 187-  7* & 21 & 35 & 19.123 & +46 & 33 & 41.32 & 1952.7 & 17.2 & 0.93 & 0.459 & 200 & 15.3 & 25 & 0.6 & WD \\
70 & 697- 45 & 21 & 31 & 21.313 & $-$5 & 11 & 16.35 & 1954.5 & 15.0 & 1.54 & 0.374 & 96 & 11.0 & 66 & 0.2 & SD \\
71 & 16- 36 & 5 & 37 & 59.137 & +79 & 31 & 7.05 & 1955.0 & 18.9 & 0.99 & 1.192 & 143 & 15.3 & 51 & 0.6 & WD \\
72 & 877- 22 & 22 & 52 & 25.884 & $-$22 & 20 & 1.71 & 1982.8 & 13.5 & 1.70 & 0.291 & 192 & 13.9 & 8 & 0.1 & MS \\
73 & R  66 & 5 & 49 & 56.578 & +36 & 50 & 46.56 & 1954.9 & 12.5 & 1.64 & 0.510 & 165 & 12.9 & 8 & 0.2 & MS \\
74 &  & 5 & 48 & 23.867 & +7 & 45 & 50.89 & 1955.9 & 14.5 & 1.53 & 0.276 & 165 & 11.0 & 52 & 0.4 & MS \\
75 & 404-  7 & 23 & 57 & 50.184 & +19 & 48 & 54.14 & 1954.7 & 17.0 & 1.04 & 0.308 & 33 & 15.4 & 21 & 0.6 & WD \\
76 & 747- 11 & 17 & 11 & 25.904 & $-$14 & 47 & 40.78 & 1954.5 & 14.5 & 0.32 & 0.371 & 132 & 13.8 & 14 & 0.6 & WD \\
77 & 382- 55 & 14 & 59 & 51.924 & +21 & 24 & 57.91 & 1950.3 & 17.6 & 1.92 & 0.218 & 283 & 17.8 & 9 & 0.1 & MS \\
78 & 192- 23 & 0 & 1 & 45.551 & +41 & 36 & 2.69 & 1954.8 & 14.8 & 1.53 & 0.299 & 141 & 11.0 & 60 & 0.4 & MS \\
79 & 787- 49 & 9 & 29 & 42.151 & $-$17 & 32 & 36.06 & 1954.2 & 16.0 & 0.38 & 0.447 & 134 & 14.1 & 24 & 0.6 & WD \\
80 &  & 15 & 27 & 44.993 & $-$9 & 1 & 18.36 & 1955.4 & 15.5 & 1.70 & 0.318 & 172 & 13.9 & 20 & 0.1 & MS \\
81 &  & 5 & 44 & 0.906 & +40 & 57 & 36.75 & 1953.0 & 15.8 & 0.33 & 1.229 & 147 & 13.8 & 25 & 0.6 & WD \\
82 &  & 18 & 39 & 28.561 & +4 & 11 & 48.05 & 1950.5 & 15.6 & 1.26 & 0.506 & 240 & 8.3 & 287 & 0.4 & SD \\
83 & L 560-9 & 18 & 8 & 7.463 & $-$30 & 55 & 37.12 & 1977.5 & 16.7 & 0.93 & 0.300 & 204 & 15.3 & 20 & 0.6 & WD \\
84 & 813- 32 & 19 & 57 & 26.935 & $-$17 & 30 & 16.64 & 1953.6 & 14.8 & 1.53 & 0.499 & 98 & 11.0 & 60 & 0.2 & SD \\
85 & 544- 37 & 8 & 15 & 18.924 & +4 & 55 & 46.72 & 1949.9 & 18.1 & 1.48 & 0.214 & 162 & 10.0 & 422 & 0.3 & SD \\
86 & * & 21 & 10 & 59.850 & +46 & 57 & 47.02 & 1952.5 & 14.6 & 1.26 & 0.395 & 218 & 8.3 & 181 & 0.4 & SD \\
87 & 82- 44 & 3 & 0 & 58.173 & +59 & 36 & 40.80 & 1954.1 & 15.2 & 1.59 & 0.223 & 92 & 11.9 & 44 & 0.3 & MS \\
88 & R 341 & 3 & 6 & 15.472 & +51 & 3 & 45.96 & 1953.8 & 13.3 & 1.48 & 0.846 & 124 & 10.0 & 46 & 0.3 & SD \\
89 & * & 13 & 14 & 9.166 & +6 & 18 & 43.70 & 1956.2 & 15.8 & 1.37 & 0.334 & 216 & 8.5 & 294 & 0.4 & SD \\
90 & 264- 49 & 11 & 24 & 9.491 & +35 & 47 & 30.88 & 1953.2 & 18.5 & 1.54 & 0.277 & 269 & 16.6 & 25 & 0.6 & WD \\
91 & 336-  6* & 19 & 7 & 38.155 & +32 & 31 & 45.32 & 1950.5 & 12.2 & 1.43 & 1.635 & 49 & 9.0 & 43 & 0.4 & SD \\
92 &  & 23 & 57 & 40.443 & +23 & 19 & 8.49 & 1950.6 & 12.3 & 1.42 & 1.460 & 135 & 9.0 & 45 & 0.4 & SD \\
93 &  & 18 & 47 & 15.693 & $-$17 & 25 & 57.59 & 1954.5 & 14.5 & 1.32 & 0.480 & 201 & 8.4 & 167 & 0.4 & SD \\
94 & 372-  4 & 10 & 20 & 42.745 & +20 & 27 & 58.46 & 1955.3 & 19.0 & 0.71 & 0.265 & 202 & 15.0 & 61 & 0.6 & WD \\
95 & R  28 & 4 & 13 & 0.374 & +52 & 37 & 18.86 & 1954.8 & 13.7 & 1.65 & 0.910 & 203 & 12.9 & 14 & 0.2 & MS \\
96 &  & 4 & 48 & 8.475 & +48 & 32 & 33.90 & 1953.8 & 16.5 & 1.32 & 0.503 & 123 & 15.8 & 14 & 0.6 & WD \\
97 &  & 23 & 9 & 36.225 & +33 & 12 & 40.10 & 1954.7 & 13.1 & 1.48 & 0.366 & 75 & 10.0 & 42 & 0.5 & MS \\
98 & 642- 53* & 23 & 21 & 16.024 & +1 & 2 & 36.62 & 1953.8 & 19.1 & 0.82 & 0.267 & 207 & 15.2 & 60 & 0.6 & WD \\
99 & 642- 52 & 23 & 21 & 16.024 & +1 & 2 & 36.62 & 1953.8 & 19.1 & 0.82 & 0.267 & 207 & 15.2 & 60 & 0.6 & WD \\
100 & 66- 58 & 13 & 32 & 49.849 & +65 & 51 & 28.96 & 1955.4 & 15.6 & 1.59 & 0.254 & 334 & 11.9 & 53 & 0.3 & MS \\

\tablebreak
101 & 809- 20 & 18 & 15 & 13.220 & $-$19 & 23 & 41.54 & 1954.5 & 12.8 & 0.93 & 0.382 & 162 & 7.6 & 112 & 0.5 & SD \\
102 &  & 15 & 24 & 38.144 & $-$6 & 49 & 7.71 & 1955.4 & 15.9 & 1.70 & 0.477 & 212 & 13.9 & 25 & 0.1 & MS \\
103 & 152- 27 & 15 & 29 & 29.620 & $-$61 & 46 & 28.80 & 1980.3 & 16.2 & 0.49 & 0.210 & 189 & 14.5 & 22 & 0.6 & WD \\
104 & 358-663 & 4 & 18 & 24.239 & +22 & 11 & 51.22 & 1950.9 & 18.0 & 1.54 & 0.404 & 136 & 16.6 & 20 & 0.6 & WD \\
105 &  & 14 & 55 & 2.824 & +43 & 1 & 45.61 & 1955.2 & 11.9 & 1.15 & 0.302 & 207 & 8.0 & 59 & 0.5 & SD \\
106 & 700- 35 & 22 & 32 & 47.803 & $-$5 & 57 & 10.13 & 1954.5 & 11.8 & 1.10 & 0.266 & 242 & 7.9 & 61 & 0.5 & SD \\
107 & 241- 23 & 0 & 31 & 3.915 & +36 & 40 & 50.36 & 1951.8 & 15.8 & 1.48 & 0.257 & 174 & 10.0 & 146 & 0.3 & SD \\
108 & 555- 21 & 12 & 27 & 54.387 & +5 & 12 & 33.92 & 1956.2 & 14.1 & 1.70 & 0.576 & 244 & 13.9 & 11 & 0.1 & MS \\
109 & 838- 25 & 6 & 14 & 16.133 & $-$23 & 10 & 16.91 & 1982.6 & 15.1 & 0.98 & 0.388 & 156 & 7.7 & 298 & 0.5 & SD \\
110 &  & 17 & 46 & 23.657 & $-$18 & 6 & 57.04 & 1950.5 & 13.7 & 1.65 & 0.210 & 207 & 12.9 & 14 & 0.2 & MS \\
111 & 29- 23 & 0 & 43 & 57.017 & +75 & 12 & 26.58 & 1954.7 & 18.3 & 0.82 & 0.302 & 104 & 15.2 & 41 & 0.6 & WD \\
112 & 782- 13 & 7 & 19 & 21.571 & $-$19 & 4 & 53.58 & 1953.0 & 13.2 & 1.75 & 0.192 & 12 & 14.9 & 5 & 0.1 & MS \\
113 & 152- 10 & 1 & 38 & 30.185 & +47 & 32 & 24.66 & 1953.8 & 18.3 & 1.26 & 0.362 & 107 & 15.7 & 33 & 0.6 & WD \\
114 & 458- 12 & 21 & 36 & 9.975 & +19 & 5 & 7.41 & 1951.7 & 11.9 & 1.04 & 0.326 & 76 & 7.8 & 66 & 0.5 & SD \\
115 &  & 23 & 10 & 3.421 & +63 & 58 & 15.31 & 1952.6 & 14.7 & 0.38 & 0.400 & 175 & 14.1 & 13 & 0.6 & WD \\
116 &  & 23 & 15 & 24.162 & +9 & 44 & 42.71 & 1951.6 & 13.7 & 1.21 & 0.412 & 74 & 8.2 & 130 & 0.5 & SD \\
117 & - 3:5711* & 23 & 49 & 22.223 & $-$2 & 34 & 26.83 & 1954.6 & 10.8 & 0.67 & 0.260 & 93 & 6.6 & 72 & 0.6 & SD \\
118 & 377- 12 & 12 & 33 & 50.545 & +22 & 34 & 32.05 & 1955.4 & 18.6 & 1.59 & 0.307 & 260 & 16.9 & 22 & 0.6 & WD \\
119 & 244-  7 & 1 & 48 & 42.049 & +38 & 16 & 21.41 & 1954.7 & 13.9 & 1.75 & 0.276 & 127 & 14.9 & 6 & 0.1 & MS \\
120 & 763-  7* & 23 & 42 & 19.220 & $-$13 & 56 & 29.82 & 1953.6 & 15.4 & 1.37 & 0.295 & 57 & 8.5 & 244 & 0.4 & SD \\
121 &  & 17 & 30 & 20.955 & +19 & 12 & 37.12 & 1951.5 & 13.5 & 1.48 & 0.396 & 104 & 10.0 & 51 & 0.3 & SD \\
122 & R 600 & 4 & 41 & 20.420 & +22 & 54 & 52.55 & 1950.9 & 13.0 & 1.32 & 0.650 & 145 & 8.4 & 84 & 0.4 & SD \\
123 &  & 6 & 59 & 29.457 & +19 & 30 & 43.79 & 1951.8 & 13.3 & 1.48 & 0.280 & 225 & 10.0 & 46 & 0.5 & MS \\
124 & 629- 12 & 18 & 6 & 21.109 & +2 & 3 & 21.23 & 1953.5 & 11.9 & 0.87 & 0.358 & 210 & 7.5 & 77 & 0.5 & SD \\
125 & 552- 14 & 11 & 11 & 55.685 & +3 & 37 & 32.06 & 1955.3 & 18.2 & 0.66 & 0.377 & 255 & 15.0 & 46 & 0.6 & WD \\
126 & 173- 45 & 13 & 41 & 50.645 & +47 & 0 & 7.27 & 1956.3 & 17.3 & 1.70 & 0.201 & 282 & 13.9 & 47 & 0.1 & MS \\
127 & 285-  9 & 21 & 12 & 29.944 & +35 & 55 & 58.31 & 1951.5 & 13.2 & 1.75 & 0.181 & 70 & 14.9 & 5 & 0.1 & MS \\
128 & 757-135 & 21 & 13 & 8.753 & $-$9 & 48 & 56.69 & 1953.7 & 17.3 & 1.75 & 0.181 & 216 & 14.9 & 30 & 0.1 & MS \\
129 & 753-  7 & 19 & 36 & 8.097 & $-$11 & 40 & 39.10 & 1951.6 & 17.9 & 1.31 & 0.248 & 186 & 15.8 & 26 & 0.6 & WD \\
130 &  & 0 & 41 & 57.203 & +57 & 48 & 4.82 & 1952.7 & 14.0 & 1.64 & 0.235 & 106 & 12.9 & 17 & 0.2 & MS \\
131 & 80 -81 & 1 & 43 & 23.245 & +62 & 39 & 32.99 & 1954.8 & 15.4 & 1.59 & 0.184 & 156 & 11.9 & 49 & 0.3 & MS \\
132 &  & 2 & 8 & 43.243 & +25 & 36 & 23.04 & 1953.8 & 14.1 & 1.48 & 0.332 & 79 & 10.0 & 67 & 0.3 & SD \\
133 & 196- 61 & 2 & 25 & 29.837 & +44 & 47 & 46.24 & 1952.0 & 12.4 & 0.99 & 0.204 & 238 & 7.7 & 86 & 0.5 & SD \\
134 & 711- 11 & 2 & 50 & 15.703 & $-$12 & 19 & 8.38 & 1955.9 & 16.2 & 1.70 & 0.226 & 84 & 13.9 & 28 & 0.1 & MS \\
135 & 119- 44 & 5 & 28 & 3.987 & +54 & 55 & 40.68 & 1955.0 & 15.7 & 1.59 & 0.206 & 142 & 11.9 & 56 & 0.3 & MS \\
136 & 58-151 & 7 & 20 & 3.997 & +68 & 27 & 48.50 & 1953.2 & 18.7 & 1.70 & 0.220 & 223 & 13.9 & 89 & 0.1 & SD \\
137 & 256- 19 & 7 & 30 & 9.144 & +32 & 48 & 32.43 & 1953.1 & 18.5 & 1.59 & 0.226 & 249 & 11.9 & 203 & 0.2 & SD \\
138 & 678- 54 & 13 & 42 & 12.158 & $-$5 & 59 & 1.36 & 1952.4 & 19.0 & 1.04 & 0.235 & 234 & 15.4 & 53 & 0.6 & WD \\
139 & 679- 21 & 14 & 4 & 49.451 & $-$5 & 31 & 20.97 & 1957.3 & 17.9 & 0.38 & 0.244 & 253 & 14.1 & 56 & 0.6 & WD \\
140 & 329- 55 & 16 & 20 & 36.017 & +29 & 15 & 18.04 & 1954.5 & 16.4 & 1.48 & 0.285 & 184 & 10.0 & 193 & 0.3 & SD \\
141 &  & 16 & 50 & 22.992 & $-$1 & 46 & 17.72 & 1950.5 & 13.8 & 0.93 & 0.257 & 198 & 7.6 & 178 & 0.5 & SD \\
142 &  & 19 & 31 & 30.369 & +32 & 20 & 51.08 & 1951.5 & 14.9 & 1.48 & 0.310 & 341 & 10.0 & 97 & 0.3 & SD \\
143 & 338-  2 & 19 & 50 & 1.235 & +32 & 34 & 51.31 & 1953.5 & 12.3 & 0.76 & 0.526 & 62 & 7.1 & 110 & 0.6 & SD \\
144 & 105-523 & 20 & 3 & 9.374 & +61 & 2 & 39.18 & 1952.6 & 12.7 & 1.26 & 0.187 & 13 & 7.9 & 88 & 0.7 & MS \\
145 & 816- 34 & 21 & 0 & 36.809 & $-$18 & 16 & 44.59 & 1954.6 & 17.5 & 1.37 & 0.198 & 207 & 8.5 & 643 & 0.4 & SD \\
146 & 48-526 & 22 & 8 & 58.979 & +70 & 41 & 41.01 & 1952.6 & 18.6 & 1.04 & 0.247 & 44 & 15.4 & 44 & 0.6 & WD \\

\enddata
\tablenotetext{}{Events are given in the order of increasing $\tau$ (see Table 5). Names are taken from NLTT. Right ascension and declination come from USNO-A2.0. Equinox J2000, epoch is that of the plate and is given in a separate column. Visual magnitude is in Tycho system, transformed from USNO-A2.0 photographic magnitudes. Color is in Johnson system, also transformed from photographic magnitudes. Proper motions are from NLTT, but in equinox J2000. For distance, physical parameters and class, see \S\ 3.1 (MS - main sequence star, SD - subdwarf, WD - white dwarf).}
\normalsize
\end{deluxetable}

\def\kms{{\rm km}\,{\rm s}^{-1}}
\begin{deluxetable}{l r r r r r r r r r r r r r} 
 \tablecaption{NLTT - source star and event properties}
\scriptsize
\tablewidth{0pt}
\tablenum{5}
 \tablehead{
   \colhead{Event}      &
   \multicolumn{3}{c}{RA}  &
   \multicolumn{3}{c}{DEC}  &
   \colhead{$V$} &
   \colhead{$B$$-$$V$} &
   \colhead{$\tau$} &
   \colhead{$\tau^*$} &
   \colhead{$d_{2000}$} &
   \colhead{$t_0$} &
   \colhead{$\beta$} \\
   \colhead{\#} &
   \colhead{h} &
   \colhead{m} &
   \colhead{s} &
   \colhead{$\circ$} &
   \colhead{$'$} &
   \colhead{$''$} &
   \colhead{} &
   \colhead{} &
   \colhead{hr} &
   \colhead{hr} &
   \colhead{$''$} &
   \colhead{yr} &
   \colhead{mas} \\
 }
\startdata
1 & 0 & 35 & 54.579 & +52 & 41 & 10.80 & 15.8 & 0.76 & 0.8 & 0.8 & 11.7 & 2014.8 & 292 \\
2 & 19 & 21 & 38.138 & +20 & 51 & 49.73 & 18.4 & 1.81 & 2.3 & 2.3 & 15.5 & 2008.9 & 167 \\
3 & 23 & 6 & 22.538 & +65 & 3 & 28.89 & 16.2 & 2.14 & 3.2 & 3.2 & 3.1 & 2009.5 & 40 \\
4 & 20 & 27 & 29.604 & $-$13 & 17 & 52.59 & 17.6 & 0.44 & 4.0 & 0.0 & 2.4 & 2006.5 & 97 \\
5 & 2 & 31 & 56.861 & $-$8 & 32 & 7.50 & 15.3 & 0.98 & 5.8 & 0.0 & 4.3 & 2014.2 & 62 \\
6 & 19 & 56 & 29.051 & $-$1 & 2 & 42.39 & 15.3 & 0.93 & 6.5 & 0.1 & 11.7 & 2014.7 & 1845 \\
7 & 16 & 35 & 13.806 & +35 & 47 & 19.73 & 19.4 & 1.21 & 6.9 & 6.9 & 2.5 & 2011.2 & 244 \\
8 & 11 & 24 & 12.471 & +21 & 21 & 39.64 & 14.1 & 0.99 & 9.4 & 0.6 & 7.2 & 2006.2 & 3094 \\
9 & 7 & 50 & 15.485 & +7 & 11 & 15.24 & 17.0 & 0.71 & 18.0 & 3.5 & 23.0 & 2012.9 & 677 \\
10 & 23 & 18 & 6.898 & +49 & 28 & 12.46 & 14.8 & 0.49 & 26.5 & 26.5 & 1.7 & 2005.2 & 201 \\
11 & 8 & 11 & 58.445 & +8 & 45 & 29.77 & 18.7 & 0.87 & 28.2 & 3.3 & 54.6 & 2010.5 & 2739 \\
12 & 17 & 36 & 11.179 & +23 & 48 & 22.91 & 18.4 & 0.93 & 28.7 & 0.0 & 1.2 & 2006.3 & 139 \\
13 & 23 & 41 & 54.678 & +44 & 10 & 17.02 & 17.8 & 1.21 & 43.3 & 7.0 & 21.1 & 2012.8 & 3970 \\
14 & 1 & 9 & 3.653 & $-$10 & 42 & 13.89 & 19.0 & 0.77 & 49.0 & 49.0 & 2.4 & 2012.1 & 196 \\
15 & 21 & 0 & 36.351 & $-$18 & 16 & 53.75 & 15.9 & 0.27 & 52.8 & 0.0 & 1.3 & 2005.7 & 699 \\
16 & 16 & 6 & 35.790 & +24 & 28 & 36.69 & 19.4 & 0.54 & 73.2 & 73.2 & 4.7 & 2014.9 & 29 \\
17 & 5 & 50 & 24.507 & +17 & 19 & 11.32 & 18.6 & 1.04 & 73.6 & 12.3 & 8.6 & 2014.6 & 735 \\
18 & 2 & 25 & 41.586 & +42 & 27 & 5.97 & 18.6 & 1.21 & 74.0 & 74.0 & 2.4 & 2010.3 & 136 \\
19 & 12 & 38 & 41.040 & $-$19 & 21 & 28.98 & 19.1 & 1.59 & 117.7 & 117.7 & 4.1 & 2011.6 & 26 \\
20 & 19 & 38 & 48.723 & +35 & 12 & 44.22 & 15.7 & 1.48 & 120.4 & 120.4 & 7.7 & 2009.8 & 58 \\
21 & 16 & 1 & 48.019 & +30 & 30 & 44.81 & 16.9 & 0.44 & 121.0 & 0.0 & 2.3 & 2010.4 & 504 \\
22 & 17 & 55 & 48.684 & $-$7 & 36 & 1.16 & 18.2 & 1.32 & 128.7 & 128.7 & 3.2 & 2012.6 & 113 \\
23 & 16 & 34 & 18.489 & +57 & 9 & 59.07 & 17.0 & 0.66 & 150.1 & 35.0 & 21.4 & 2013.2 & 621 \\
24 & 1 & 47 & 55.998 & +60 & 7 & 28.82 & 17.3 & 0.60 & 161.5 & 24.9 & 1.5 & 2006.0 & 569 \\
25 & 17 & 42 & 10.236 & $-$8 & 49 & 4.78 & 17.9 & 1.48 & 162.7 & 41.9 & 9.3 & 2009.6 & 591 \\
26 & 17 & 19 & 3.782 & +28 & 5 & 3.45 & 17.1 & 0.66 & 163.7 & 163.7 & 1.7 & 2007.1 & 7 \\
27 & 13 & 8 & 25.655 & +12 & 26 & 36.56 & 19.2 & 0.55 & 171.9 & 171.9 & 1.8 & 2006.3 & 45 \\
28 & 21 & 40 & 29.996 & +54 & 0 & 33.33 & 16.7 & 0.93 & 178.8 & 58.1 & 4.6 & 2011.0 & 526 \\
29 & 4 & 12 & 58.173 & +52 & 36 & 35.13 & 16.6 & 0.93 & 180.3 & 15.7 & 7.0 & 2007.6 & 1018 \\
30 & 18 & 6 & 31.383 & $-$30 & 9 & 52.44 & 17.7 & 0.71 & 219.3 & 44.3 & 1.6 & 2006.0 & 537 \\
31 & 17 & 15 & 23.471 & +1 & 19 & 23.59 & 16.7 & 0.76 & 236.5 & 236.5 & 4.3 & 2011.7 & 159 \\
32 & 0 & 44 & 1.323 & +75 & 12 & 21.45 & 16.3 & 1.32 & 252.8 & 0.0 & 3.7 & 2011.8 & 927 \\
33 & 8 & 31 & 8.473 & $-$20 & 42 & 4.44 & 17.4 & 1.32 & 270.0 & 270.0 & 1.9 & 2007.5 & 256 \\
34 & 20 & 34 & 33.169 & +7 & 57 & 34.37 & 15.9 & 1.21 & 290.9 & 10.0 & 2.1 & 2005.0 & 1047 \\
35 & 7 & 13 & 41.229 & $-$13 & 28 & 7.39 & 18.9 & 0.71 & 302.3 & 20.1 & 14.6 & 2011.2 & 2623 \\
36 & 5 & 10 & 31.270 & +31 & 17 & 30.97 & 19.3 & 0.82 & 333.5 & 31.4 & 4.0 & 2005.7 & 958 \\
37 & 23 & 5 & 16.976 & +71 & 23 & 13.35 & 17.8 & 0.77 & 335.4 & 0.0 & 2.8 & 2010.1 & 425 \\
38 & 20 & 43 & 21.091 & +55 & 21 & 11.31 & 19.2 & 0.76 & 340.5 & 65.8 & 23.9 & 2012.3 & 3761 \\
39 & 17 & 7 & 15.943 & +19 & 25 & 42.30 & 18.5 & 0.82 & 367.5 & 58.2 & 1.9 & 2010.4 & 496 \\
40 & 7 & 11 & 12.028 & +43 & 29 & 50.10 & 17.0 & 1.37 & 377.0 & 377.0 & 9.7 & 2014.2 & 241 \\
41 & 19 & 22 & 1.631 & +7 & 2 & 24.64 & 16.0 & 1.54 & 381.2 & 18.5 & 10.1 & 2011.4 & 3233 \\
42 & 22 & 36 & 36.875 & +53 & 3 & 6.69 & 17.7 & 0.66 & 405.2 & 334.9 & 1.9 & 2007.3 & 330 \\
43 & 9 & 9 & 55.868 & $-$11 & 26 & 19.66 & 19.2 & 0.49 & 432.9 & 432.9 & 6.3 & 2013.1 & 178 \\
44 & 3 & 43 & 54.491 & +63 & 39 & 48.85 & 18.4 & 1.37 & 451.3 & 451.3 & 9.1 & 2009.5 & 94 \\
45 & 4 & 31 & 12.144 & +58 & 58 & 47.85 & 17.6 & 0.76 & 568.5 & 197.6 & 23.1 & 2009.2 & 7577 \\
46 & 0 & 9 & 53.777 & +53 & 1 & 15.84 & 18.7 & 0.82 & 587.7 & 0.5 & 2.9 & 2009.5 & 1790 \\
47 & 20 & 14 & 44.986 & +61 & 46 & 37.82 & 18.9 & 0.93 & 603.9 & 12.2 & 7.3 & 2010.2 & 1608 \\
48 & 20 & 33 & 59.518 & +64 & 19 & 8.75 & 15.4 & 1.32 & 613.7 & 34.9 & 3.1 & 2006.8 & 978 \\
49 & 2 & 12 & 0.119 & +32 & 21 & 43.15 & 17.6 & 0.82 & 628.7 & 31.4 & 5.7 & 2009.9 & 1126 \\
50 & 2 & 7 & 4.257 & +49 & 38 & 36.99 & 18.5 & 0.10 & 631.3 & 631.3 & 7.2 & 2014.4 & 245 \\

\tablebreak
51 & 22 & 1 & 8.543 & +29 & 9 & 37.00 & 17.4 & 1.48 & 655.9 & 3.7 & 5.8 & 2009.1 & 2063 \\
52 & 10 & 56 & 24.791 & +7 & 0 & 27.92 & 15.8 & 0.76 & 658.7 & 421.7 & 68.5 & 2014.0 & 19571 \\
53 & 1 & 4 & 5.115 & +59 & 38 & 0.75 & 18.4 & 0.93 & 720.1 & 298.3 & 3.4 & 2008.1 & 466 \\
54 & 17 & 18 & 47.071 & $-$29 & 46 & 5.85 & 16.2 & 1.43 & 760.1 & 1.6 & 3.5 & 2010.5 & 2424 \\
55 & 1 & 48 & 49.863 & +55 & 2 & 6.08 & 17.5 & 1.26 & 771.6 & 771.6 & 3.7 & 2013.2 & 90 \\
56 & 20 & 11 & 13.029 & +16 & 11 & 11.76 & 17.3 & 0.82 & 825.3 & 256.9 & 5.4 & 2009.3 & 538 \\
57 & 19 & 38 & 32.546 & $-$2 & 51 & 17.35 & 17.7 & 0.93 & 846.7 & 80.8 & 3.6 & 2012.2 & 705 \\
58 & 16 & 34 & 41.572 & $-$9 & 1 & 54.26 & 18.4 & 1.15 & 859.0 & 845.8 & 1.5 & 2008.2 & 245 \\
59 & 18 & 2 & 31.573 & +5 & 44 & 38.70 & 14.9 & 1.43 & 898.2 & 0.2 & 4.5 & 2008.7 & 1866 \\
60 & 2 & 19 & 3.789 & +35 & 21 & 14.03 & 19.8 & 0.38 & 921.1 & 111.4 & 10.9 & 2013.7 & 863 \\
61 & 12 & 38 & 32.897 & +35 & 13 & 6.02 & 12.7 & 1.48 & 947.8 & 0.0 & 4.3 & 2014.8 & 1646 \\
62 & 0 & 28 & 54.282 & +50 & 22 & 36.51 & 18.7 & 0.71 & 1080.0 & 3.1 & 4.4 & 2008.7 & 2155 \\
63 & 12 & 21 & 50.167 & +6 & 43 & 30.54 & 18.2 & 1.59 & 1180.4 & 504.0 & 5.3 & 2007.3 & 459 \\
64 & 22 & 32 & 58.125 & +53 & 47 & 43.99 & 17.7 & 1.32 & 1593.4 & 77.9 & 15.7 & 2011.7 & 2409 \\
65 & 17 & 37 & 27.797 & +71 & 3 & 52.41 & 19.2 & 1.09 & 1695.4 & 79.0 & 6.6 & 2013.5 & 1081 \\
66 & 19 & 3 & 14.965 & $-$13 & 34 & 17.86 & 15.7 & 0.82 & 1727.0 & 220.8 & 7.6 & 2007.2 & 5161 \\
67 & 5 & 44 & 4.547 & +40 & 56 & 36.51 & 18.5 & 0.82 & 1851.0 & 57.2 & 15.4 & 2012.4 & 2254 \\
68 & 21 & 35 & 18.524 & +46 & 33 & 17.42 & 16.3 & 0.93 & 1861.0 & 8.1 & 3.7 & 2006.3 & 2263 \\
69 & 21 & 35 & 18.524 & +46 & 33 & 17.42 & 16.3 & 0.93 & 1861.0 & 8.1 & 3.7 & 2006.3 & 2263 \\
70 & 21 & 31 & 22.614 & $-$5 & 11 & 19.26 & 16.0 & 0.98 & 1863.1 & 104.7 & 2.8 & 2007.0 & 920 \\
71 & 5 & 38 & 15.352 & +79 & 30 & 15.74 & 14.7 & 1.10 & 1872.0 & 206.6 & 14.6 & 2011.8 & 3887 \\
72 & 22 & 52 & 25.820 & $-$22 & 20 & 10.76 & 18.6 & 1.15 & 1963.9 & 47.6 & 4.2 & 2013.9 & 1000 \\
73 & 5 & 49 & 57.019 & +36 & 50 & 20.82 & 18.9 & 1.21 & 2080.0 & 42.5 & 3.5 & 2006.4 & 1393 \\
74 & 5 & 48 & 24.167 & +7 & 45 & 37.30 & 17.3 & 0.49 & 2098.8 & 86.2 & 2.3 & 2007.6 & 860 \\
75 & 23 & 57 & 50.928 & +19 & 49 & 8.44 & 19.7 & 0.76 & 2242.2 & 57.3 & 3.9 & 2012.2 & 1019 \\
76 & 17 & 11 & 26.777 & $-$14 & 47 & 54.83 & 17.8 & 0.87 & 2586.4 & 4.1 & 2.8 & 2005.2 & 2057 \\
77 & 14 & 59 & 51.067 & +21 & 24 & 59.92 & 18.4 & 0.93 & 2959.1 & 180.1 & 1.5 & 2005.9 & 692 \\
78 & 0 & 1 & 46.608 & +41 & 35 & 51.44 & 13.1 & 0.76 & 3180.2 & 2.9 & 3.4 & 2008.9 & 2131 \\
79 & 9 & 29 & 43.422 & $-$17 & 32 & 55.88 & 18.9 & 0.77 & 3356.7 & 33.8 & 6.6 & 2014.3 & 1544 \\
80 & 15 & 27 & 45.232 & $-$9 & 1 & 36.89 & 18.1 & 0.60 & 3379.0 & 195.0 & 4.7 & 2014.6 & 848 \\
81 & 5 & 44 & 4.731 & +40 & 56 & 36.92 & 16.6 & 0.93 & 3418.2 & 429.3 & 16.6 & 2013.0 & 4224 \\
82 & 18 & 39 & 26.907 & +4 & 11 & 34.35 & 16.2 & 1.21 & 3428.8 & 745.7 & 3.3 & 2006.4 & 643 \\
83 & 18 & 8 & 7.261 & $-$30 & 55 & 47.57 & 17.4 & 2.36 & 3654.7 & 5.7 & 4.3 & 2012.9 & 1824 \\
84 & 19 & 57 & 29.012 & $-$17 & 30 & 19.74 & 17.4 & 0.88 & 4153.8 & 323.8 & 6.8 & 2013.5 & 931 \\
85 & 8 & 15 & 19.197 & +4 & 55 & 34.92 & 16.5 & 0.66 & 4355.4 & 0.1 & 1.8 & 2008.2 & 289 \\
86 & 21 & 10 & 58.628 & +46 & 57 & 29.33 & 16.1 & 0.76 & 4638.9 & 271.9 & 3.0 & 2007.4 & 931 \\
87 & 3 & 0 & 59.817 & +59 & 36 & 41.43 & 15.7 & 1.04 & 4766.7 & 29.8 & 2.5 & 2009.9 & 1150 \\
88 & 3 & 6 & 19.919 & +51 & 3 & 19.98 & 16.3 & 1.09 & 4919.1 & 73.7 & 10.4 & 2012.0 & 2182 \\
89 & 13 & 14 & 8.421 & +6 & 18 & 29.92 & 15.0 & 1.31 & 5696.5 & 0.0 & 3.2 & 2009.1 & 909 \\
90 & 11 & 24 & 8.201 & +35 & 47 & 31.74 & 19.7 & 0.66 & 6186.1 & 79.4 & 2.9 & 2009.8 & 1120 \\
91 & 19 & 7 & 44.041 & +32 & 32 & 48.05 & 18.2 & 0.49 & 6590.8 & 144.9 & 16.4 & 2010.0 & 2023 \\
92 & 23 & 57 & 45.331 & +23 & 18 & 4.94 & 17.3 & 1.43 & 6680.6 & 539.1 & 20.6 & 2014.0 & 2665 \\
93 & 18 & 47 & 14.940 & $-$17 & 26 & 21.29 & 14.5 & 1.92 & 6935.0 & 54.1 & 4.5 & 2008.6 & 1701 \\
94 & 10 & 20 & 42.195 & +20 & 27 & 45.69 & 14.0 & 0.93 & 7497.0 & 12.9 & 3.7 & 2010.9 & 2351 \\
95 & 4 & 12 & 58.125 & +52 & 36 & 37.44 & 18.1 & 1.48 & 7537.2 & 198.2 & 5.5 & 2005.5 & 2343 \\
96 & 4 & 48 & 11.065 & +48 & 32 & 20.65 & 17.8 & 1.32 & 7898.8 & 158.6 & 6.3 & 2011.0 & 3102 \\
97 & 23 & 9 & 37.763 & +33 & 12 & 43.90 & 18.4 & 0.93 & 8163.2 & 91.1 & 3.3 & 2008.3 & 1351 \\
98 & 23 & 21 & 15.663 & +1 & 2 & 23.85 & 18.9 & 0.77 & 9066.2 & 0.0 & 1.8 & 2005.6 & 957 \\
99 & 23 & 21 & 15.663 & +1 & 2 & 23.85 & 18.9 & 0.77 & 9066.2 & 0.0 & 1.8 & 2005.6 & 957 \\
100 & 13 & 32 & 48.813 & +65 & 51 & 40.30 & 19.0 & 0.38 & 9804.3 & 863.4 & 1.8 & 2006.5 & 682 \\

\tablebreak
101 & 18 & 15 & 13.579 & $-$19 & 24 & 0.98 & 16.9 & 1.65 & 10406.1 & 167.0 & 3.0 & 2007.0 & 1274 \\
102 & 15 & 24 & 37.078 & $-$6 & 49 & 28.27 & 14.6 & 1.15 & 11879.7 & 119.5 & 5.3 & 2009.6 & 2670 \\
103 & 15 & 29 & 29.890 & $-$61 & 46 & 35.06 & 15.4 & 1.21 & 12565.1 & 125.8 & 3.3 & 2008.4 & 2824 \\
104 & 4 & 18 & 25.230 & +22 & 11 & 33.41 & 18.4 & 1.15 & 13788.2 & 52.8 & 3.5 & 2006.4 & 2365 \\
105 & 14 & 55 & 2.393 & +43 & 1 & 29.84 & 13.7 & 1.15 & 14533.9 & 145.5 & 3.9 & 2008.9 & 2879 \\
106 & 22 & 32 & 46.989 & $-$5 & 57 & 17.76 & 18.6 & 0.32 & 14750.1 & 269.9 & 2.4 & 2008.3 & 1014 \\
107 & 0 & 31 & 4.126 & +36 & 40 & 35.67 & 15.9 & 1.15 & 16817.1 & 296.5 & 2.7 & 2009.6 & 1002 \\
108 & 12 & 27 & 52.465 & +5 & 12 & 23.48 & 16.4 & 1.59 & 17022.7 & 456.5 & 6.1 & 2008.9 & 3215 \\
109 & 6 & 14 & 16.562 & $-$23 & 10 & 27.88 & 16.4 & 0.11 & 17980.3 & 752.3 & 5.7 & 2014.6 & 1014 \\
110 & 17 & 46 & 23.181 & $-$18 & 7 & 7.06 & 17.0 & 1.37 & 18248.0 & 22.5 & 2.2 & 2007.6 & 1558 \\
111 & 0 & 44 & 1.377 & +75 & 12 & 24.33 & 17.3 & 1.04 & 18700.1 & 3.4 & 3.6 & 2010.1 & 1915 \\
112 & 7 & 19 & 21.877 & $-$19 & 4 & 43.77 & 17.0 & 1.10 & 23243.6 & 9.9 & 2.6 & 2007.8 & 2154 \\
113 & 1 & 38 & 32.012 & +47 & 32 & 16.06 & 16.8 & 0.44 & 25131.9 & 199.0 & 4.4 & 2009.6 & 2753 \\
114 & 21 & 36 & 11.207 & +19 & 5 & 12.95 & 19.2 & 0.49 & 25135.2 & 534.1 & 2.8 & 2007.8 & 1097 \\
115 & 23 & 10 & 4.171 & +63 & 57 & 52.25 & 18.9 & 1.37 & 25253.3 & 272.5 & 5.3 & 2011.2 & 2858 \\
116 & 23 & 15 & 25.657 & +9 & 44 & 46.56 & 14.4 & 0.60 & 28041.6 & 127.1 & 3.4 & 2005.7 & 2415 \\
117 & 23 & 49 & 23.228 & $-$2 & 34 & 26.57 & 19.3 & 0.93 & 28674.9 & 436.4 & 3.4 & 2012.4 & 1041 \\
118 & 12 & 33 & 49.375 & +22 & 34 & 26.10 & 16.9 & 0.93 & 29929.6 & 436.2 & 4.5 & 2010.8 & 3033 \\
119 & 1 & 48 & 43.205 & +38 & 16 & 13.21 & 19.0 & 1.21 & 35659.6 & 63.4 & 3.7 & 2012.0 & 1691 \\
120 & 23 & 42 & 20.200 & $-$13 & 56 & 22.21 & 15.1 & 1.42 & 38018.7 & 0.0 & 2.8 & 2008.2 & 1397 \\
121 & 17 & 30 & 22.585 & +19 & 12 & 29.43 & 16.9 & 0.71 & 41301.9 & 89.9 & 5.4 & 2012.8 & 2002 \\
122 & 4 & 41 & 21.804 & +22 & 54 & 21.73 & 18.3 & 1.21 & 41411.3 & 458.4 & 4.7 & 2006.7 & 1835 \\
123 & 6 & 59 & 28.537 & +19 & 30 & 33.30 & 18.1 & 1.10 & 46083.0 & 83.9 & 3.6 & 2011.2 & 1696 \\
124 & 18 & 6 & 20.297 & +2 & 3 & 3.82 & 17.3 & 1.21 & 58202.4 & 108.9 & 4.9 & 2012.6 & 1947 \\
125 & 11 & 11 & 54.241 & +3 & 37 & 28.13 & 19.4 & 1.21 & 59445.1 & 200.6 & 5.3 & 2013.4 & 1775 \\
126 & 13 & 41 & 49.532 & +47 & 0 & 8.19 & 15.6 & 1.48 & 62787.4 & 0.0 & 2.9 & 2012.7 & 1434 \\
127 & 21 & 12 & 30.703 & +35 & 56 & 4.50 & 17.3 & 1.09 & 76812.7 & 405.5 & 3.3 & 2011.1 & 2621 \\
128 & 21 & 13 & 8.389 & $-$9 & 49 & 6.11 & 17.2 & 1.04 & 78326.6 & 0.0 & 2.7 & 2013.3 & 1148 \\
129 & 19 & 36 & 8.131 & $-$11 & 40 & 53.07 & 19.4 & 0.82 & 79304.8 & 62.9 & 2.6 & 2007.4 & 1884 \\
130 & 0 & 41 & 58.855 & +57 & 48 & 3.19 & 19.3 & 0.77 & $>$100000 & 245.6 & 2.9 & 2008.6 & 2092 \\
131 & 1 & 43 & 23.601 & +62 & 39 & 23.63 & 17.0 & 0.93 & $>$100000 & 121.4 & 2.0 & 2006.7 & 1525 \\
132 & 2 & 8 & 44.489 & +25 & 36 & 28.48 & 17.8 & 0.66 & $>$100000 & 594.0 & 3.1 & 2006.7 & 2175 \\
133 & 2 & 25 & 29.051 & +44 & 47 & 38.67 & 18.8 & 0.77 & $>$100000 & 593.9 & 2.4 & 2006.4 & 2034 \\
134 & 2 & 50 & 16.477 & $-$12 & 19 & 8.52 & 18.2 & 0.38 & $>$100000 & 389.9 & 1.8 & 2005.7 & 1288 \\
135 & 5 & 28 & 4.644 & +54 & 55 & 30.76 & 18.1 & 1.04 & $>$100000 & 430.1 & 2.6 & 2009.9 & 1554 \\
136 & 7 & 20 & 2.327 & +68 & 27 & 40.62 & 18.0 & 0.60 & $>$100000 & 0.1 & 2.2 & 2007.9 & 1411 \\
137 & 7 & 30 & 8.141 & +32 & 48 & 28.61 & 17.3 & 1.43 & $>$100000 & 0.1 & 2.7 & 2011.5 & 899 \\
138 & 13 & 42 & 11.445 & $-$5 & 59 & 11.55 & 18.6 & 0.27 & $>$100000 & 35.6 & 3.9 & 2014.6 & 1959 \\
139 & 14 & 4 & 48.737 & $-$5 & 31 & 26.82 & 17.3 & 1.15 & $>$100000 & 427.1 & 2.9 & 2006.2 & 2447 \\
140 & 16 & 20 & 36.102 & +29 & 15 & 0.76 & 14.1 & 0.82 & $>$100000 & 206.5 & 4.8 & 2014.7 & 2221 \\
141 & 16 & 50 & 22.520 & $-$1 & 46 & 32.75 & 16.4 & 0.71 & $>$100000 & 392.7 & 4.3 & 2014.6 & 2161 \\
142 & 19 & 31 & 30.049 & +32 & 21 & 8.21 & 16.8 & 0.99 & $>$100000 & 293.5 & 3.1 & 2008.0 & 1844 \\
143 & 19 & 50 & 3.063 & +32 & 35 & 6.21 & 18.3 & 0.87 & $>$100000 & 757.4 & 3.7 & 2005.6 & 2174 \\
144 & 20 & 3 & 9.483 & +61 & 2 & 49.76 & 19.0 & 0.87 & $>$100000 & 403.4 & 2.2 & 2008.8 & 1521 \\
145 & 21 & 0 & 36.351 & $-$18 & 16 & 53.75 & 15.9 & 0.27 & $>$100000 & 0.0 & 2.7 & 2010.7 & 1695 \\
146 & 22 & 9 & 0.546 & +70 & 41 & 52.27 & 17.9 & 1.09 & $>$100000 & 66.8 & 2.8 & 2007.3 & 2144 \\

\enddata
\tablenotetext{}{Numeration follows the numbers in table 4. Source star's right ascension and declination are at plate epoch, equinox J2000. Visual magnitude is in Tycho, and color in Johnson system (see \S\ 3.1). Event is described by $\tau$ ({\it SIM} observing time), $\tau^*$ ({\it SIM} observing time with reduced impact parameter), $d_{2000}$ lens-source separation in year 2000.0, $t_0$ time of closest approach and $\beta$, the impact parameter.}
\normalsize
\end{deluxetable}


\clearpage 


\begin{references} 
\reference{}Boden, A.\ F., Shao, M. \& Van Buren, D.\ 1998, \apj, 502, 538
\reference{hiptyc} ESA,\ 1997, The Hipparcos and Tycho Catalogues, SP-1200
\reference{flynn} Flynn, C., Sommer-Larsen, J., Fuchs, B., Graff, D.\ S. \& Salim, S.\ 1999, (astro-ph/9912264)
\reference{g1} Gould, A.\ 2000, \apj, 532, 000 (astro-ph/9905120)
\reference{luy} Luyten, W.\ J.\ 1979, 1980, New Luyten Catalogue of Stars with
 Proper Motions Larger than Two Tenths of an Arcsecond (Minneapolis: University
 of Minnesota Press)
\reference{1st} Luyten, W.\ J. \& Hughes, H.\ S.\ 1980, Proper Motion Survey with
 the Forty-Eight Inch Schmidt Telescope. LV. First Supplement to the NLTT 
Catalogue (Minneapolis: University of Minnesota) 
\reference{me} Miralda-Escud\'e, J. 1996, \apj, 470, L113
\reference{monet} Monet, D.\ 1998, BAAS, 193, 120.03
\reference{mm} Morrison, J. \& McLean, B.\ 1999, AAS/Division of Dynamical Astronomy Meeting, 31, 0510
\reference{pac} Paczy\'nski, B.\ 1995, Acta Astron., 45, 345
\reference{pac2} Paczy\'nski, B.\ 1998, \apj, 494, L23
\reference{reid} Reid, N.\ 1990, \mnras, 247, 70
\reference{refsdal} Refsdal, S.\ 1964, \mnras, 128, 295
\reference{ac2000} Urban S.\ E., Corbin T.\ E. \& Wycoff G.\ L.\ Martin, J.\ C.\, Jackson, E.\ S.\, Zacharias, M.\ I.\ 1998a, \aj, 115, 1212
\reference{act} Urban, S.\ E., Corbin, T.\ E. \& Wycoff, G.\ L.\ 1998b, \aj, 115, 2161
\end{references}
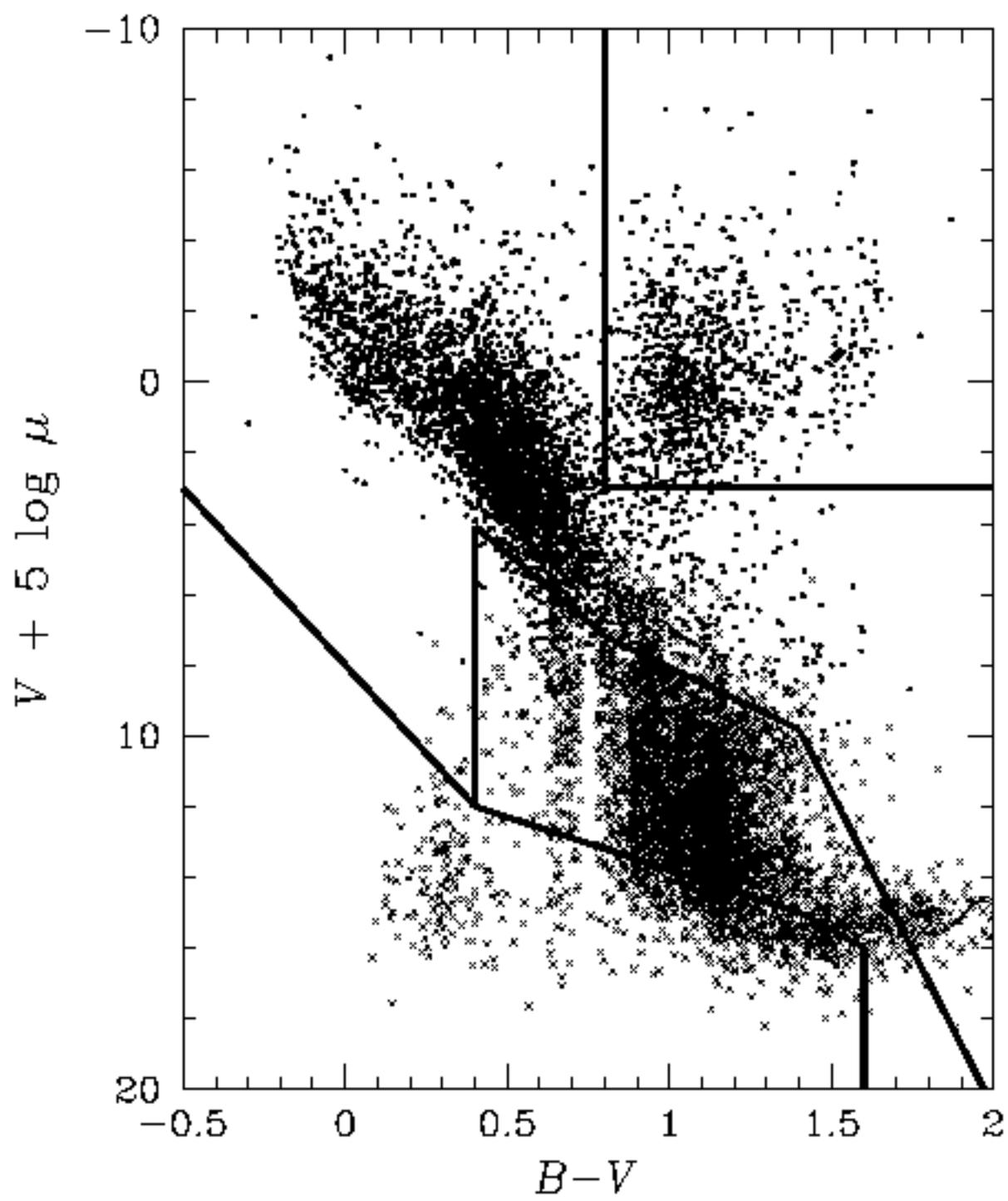
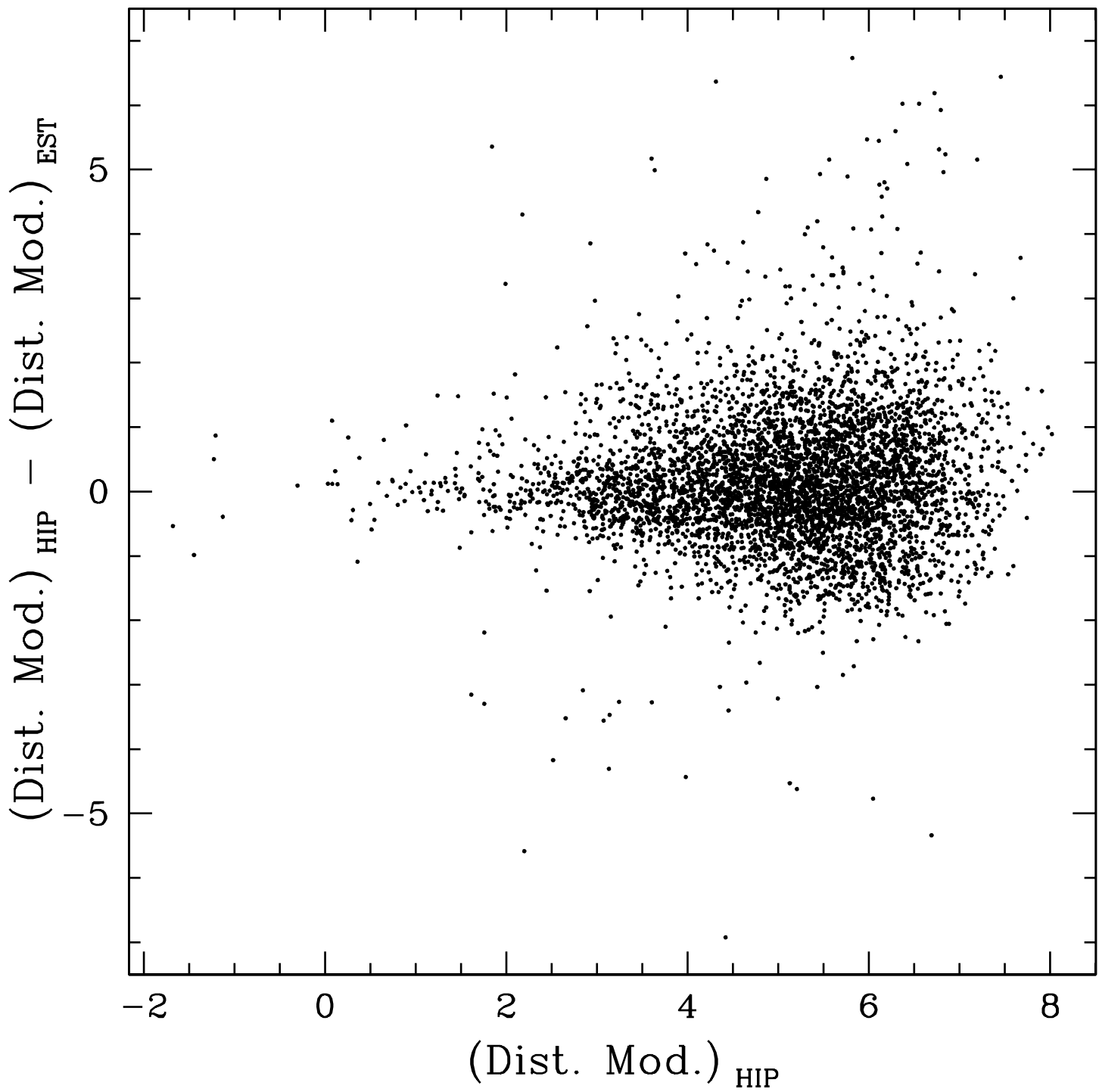
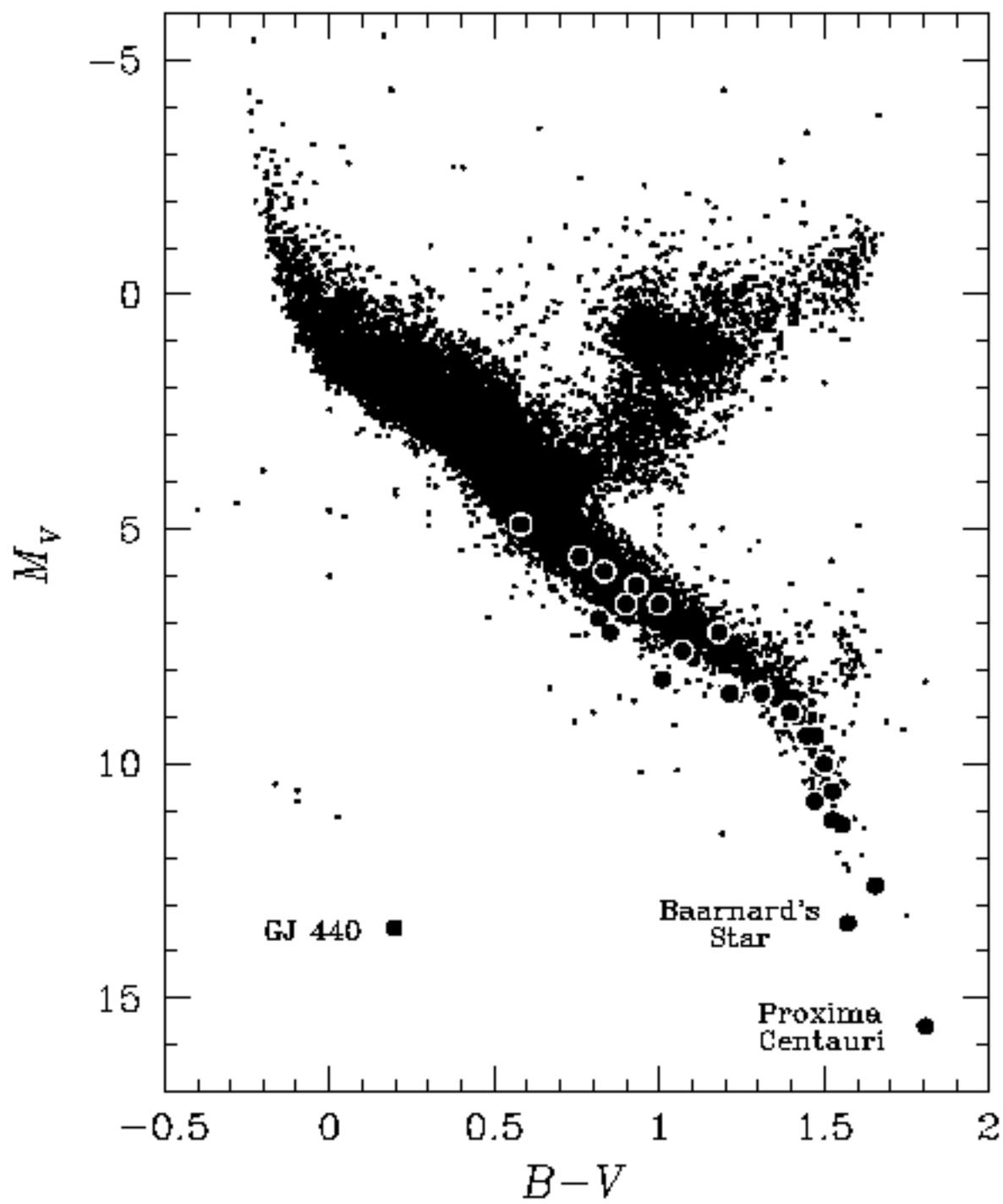
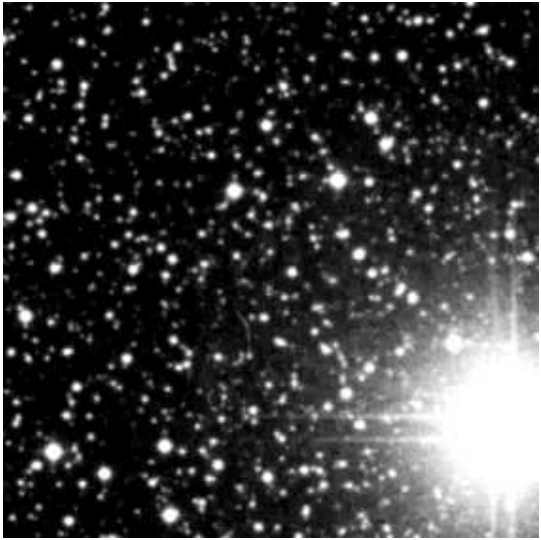
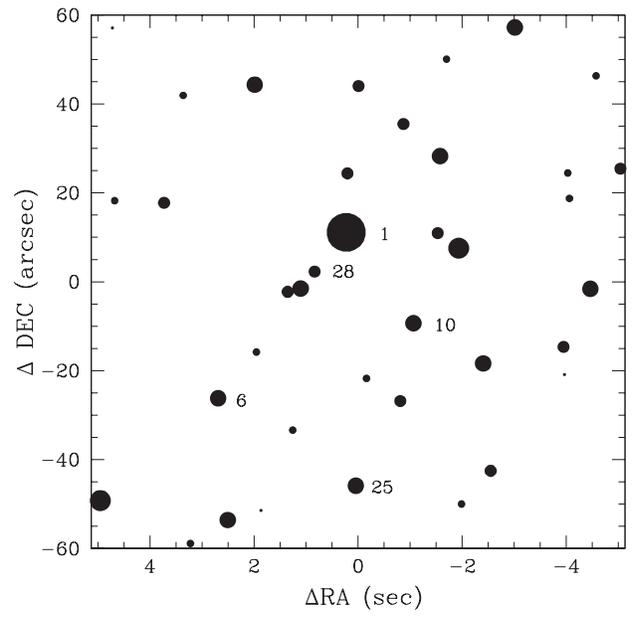
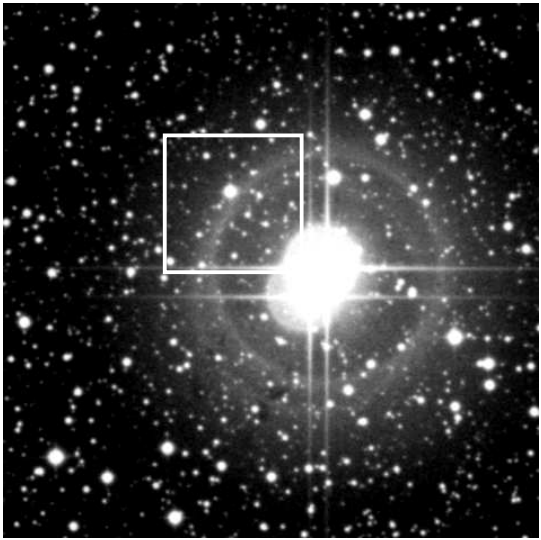
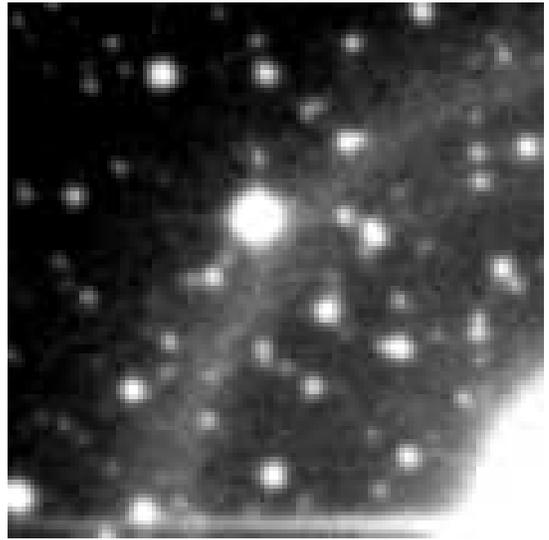
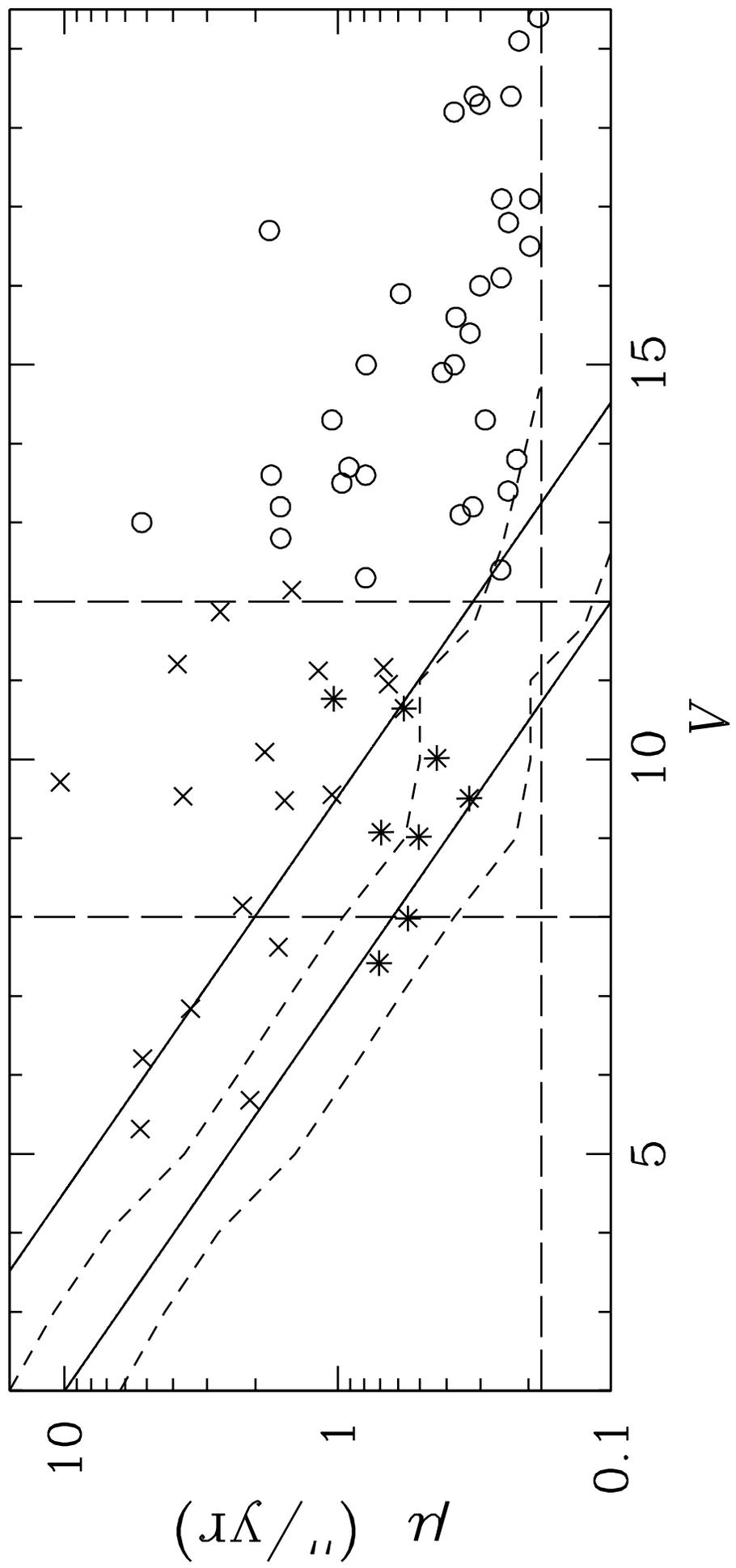
\end{document}